\documentclass{JFM-FLM_Au}

\usepackage{graphicx}
\usepackage{amsmath,amssymb}
\usepackage{bm}
\usepackage{multirow}
\usepackage{siunitx}
\usepackage{verbatim}

\lefttitle{Y. Liu \& C. R. Constante-Amores }
\righttitle{Journal of Fluid Mechanics}

\title{
Relative periodic orbits in spatially extended Kolmogorov turbulence: from global to localized recurrence
}

\author{Yaning Liu\aff{1,2} \and C. Ricardo Constante-Amores\aff{1} 
}

\affiliation{\aff{1}Department of Mechanical Science and Engineering, University of Illinois at Urbana-Champaign,  USA
\aff{2}Xingjian College, Tsinghua University, Beijing , China}

\corresau{C. Ricardo Constante-Amores, \email{crconsta@illinois.edu}}

\begin{document}

\maketitle

\begin{abstract}

Recurrent invariant solutions provide a dynamical framework for interpreting turbulence, but their computation becomes increasingly difficult in spatially extended flows, where close whole-domain recurrences are rare.
We study two-dimensional Kolmogorov flow in a 
$[L_x,L_y]= [2\pi,12\pi]$ domain with $Re=10$ using a distributed reduced-order model composed of a patch-based autoencoder and a neural ordinary differential equation.
Near-recurrence candidates are refined by latent multiple shooting, decoded to physical space, screened under the full Navier-Stokes dynamics and supplied to a full-state Newton-Krylov solver.
Fourteen candidates converge to relative periodic orbits (RPOs), with periods $3.65\leq T \leq 16.45$. The catalogue contains both domain-filling states and RPOs with recurrent activity
localised to a restricted cross-stream region.  The weakly modulated surroundings generally remain finite-amplitude and distinct from the laminar solution. Selected localised RPOs persist under continuation in Reynolds number, and recurrent states can be reconverged after truncating substantial portions of their surroundings.
 In addition, the model trained at $L_y=12\pi$
generates convergent RPO seeds at $L_y=6\pi$ and $8\pi$  without retraining. These results demonstrate that distributed learned dynamics can provide useful initial conditions for exact coherent states searches in extended domains and reveal recurrent solutions with strongly localised temporal dynamics.

\end{abstract}


\section{Introduction}
\label{sec:introduction}

The dynamical-systems description of turbulence seeks to interpret a turbulent flow as a trajectory through a high-dimensional state space organized by unstable invariant solutions of the governing equations \citep{Graham_ecs}. Equilibria, travelling waves (TW), periodic orbits and relative periodic orbits (RPOs), collectively referred to as exact coherent states (ECS), provide exact realizations of recurrent dynamical processes and have been identified in a wide range of canonical flows \citep{KAWAHARA_KIDA_2001,Faisst_Eckhardt,wedin_kerswell_2004,Viswanath_2007,Gibson_2008,Kawahara_arfm,chandler_kerswell_2013}. In wall-bounded shear flows, these solutions have provided insight into the self-sustaining processes underlying turbulent motion and into the geometry of transition and turbulence in state space \citep{willis_avila_2013,budanur_jfm_2017,Graham_ecs}.
More recently, sufficiently large collections of periodic orbits have enabled turbulent dynamics to be coarse-grained in terms of visits to recurrent states and long-time statistics to be reconstructed from weighted combinations of those states \citep{Yalniz,Page_2024}. These developments provide increasingly strong support for the view that recurrent solutions form an organizing skeleton of turbulent dynamics.

The recent work of \citet{Zhigunov_Page_2026} took an important step towards spatially extended dynamics by considering Kolmogorov flow in a  domain size of $[L_x,L_y]=[2\pi,4\pi]$. Starting from an extensive library of RPOs computed in the $[L_x,L_y]=[2\pi,2\pi]$ system, they demonstrated that larger-domain ECS can be constructed from spatial combinations of small-domain states. These included  RPOs composed of an RPO adjacent to a laminar region, quasiperiodic states constructed from pairs of small-domain RPOs, and turbulent trajectories that shadow an RPO within only part of the domain. This picture is closely related to the broader interpretation of spatiotemporal chaos as a space-time mosaic assembled from recurrent local patterns \citep{Cvitanovic_2000,gudorf2020spatiotemporal}.
A complementary question is whether localised invariant solutions can be discovered directly from turbulence in a substantially extended domain, without prescribing either their spatial support or the small-domain invariant states from which they might be assembled. 
Such a search may reveal not only localized recurrent structures, but also domain-filling RPOs that do not admit an obvious interpretation as patched combinations of smaller-domain RPOs. 
Here we address this complementary problem by searching directly within turbulent data
from the extended domain.

In spatially extended systems, close whole-domain recurrences become increasingly difficult to identify because dynamically active regions can coexist with comparatively quiescent
surroundings. This behaviour is already known in extended two-dimensional Kolmogorov flow, where spatially localised temporal dynamics arise within larger global states \citep{lucas_kerswell_2014,lucas2015recurrent}.
Long recurrence periods further amplify the sensitivity of shooting methods to errors in the initial state, period and symmetry shift. 
These limitations motivate the use of alternative strategies to identify promising candidates.

Alternative approaches have sought to relax the requirement of an accurate near-recurrent initial condition. \citet{Page_Kerswell_2020} used dynamic mode decomposition  to identify signatures of unstable periodic orbits from short turbulent trajectory segments.
A different strategy is to replace the shooting problem by an optimization over an entire candidate trajectory.
Adjoint-based variational formulations construct periodic solutions by minimizing a residual defined over a closed space-time loop and can converge from initial guesses substantially less accurate than those typically required by Newton shooting \citep{Azimi_Ashtari_Schneider_2022,Parker_Schneider_2022}. \citet{Page_2024} used a fully differentiable Navier-Stokes solver for two-dimensional Kolmogorov flow, using automatic differentiation to optimize  a trajectory-dependent recurrence loss with respect to the initial condition, period and spatial shift.
These approaches avoid, or substantially relax, the requirement that a sufficiently accurate global near recurrence first be observed in the turbulent trajectory.

A complementary strategy is to reduce the dimensionality of the search itself \citep{Page_Holey_Brenner_Kerswell_2024,glueing,alec_coutte,pipe_flow,isaac}.
\citet{glueing} used an autoencoder to construct periodic loops directly in a low-dimensional latent space, providing data-driven initial guesses for variational convergence to unstable periodic orbits without requiring near recurrences in the original trajectory.
Data-driven manifold models combine autoencoders, which provide nonlinear low-dimensional coordinates for the flow, with neural ordinary differential equations that approximate the evolution on the learned manifold \citep{alec_chaos}. In minimal turbulent shear flows and chaotic falling films, recurrent trajectories identified using such reduced dynamics have been decoded to the full state space and used as initial conditions for Newton-Krylov convergence to ECS \citep{alec_coutte,pipe_flow,isaac}.  The reduced model therefore provides a lower-dimensional search space in which candidate trajectories can be refined before expensive full-state convergence is attempted.

In this work, we search for and characterize RPOs in two-dimensional Kolmogorov flow on a spatially extended domain, $[L_x,L_y]=[2\pi,12\pi]$. We examine whether latent-space refinement can enable full-state convergence from near recurrences in this large domain, and then use the converged RPOs to investigate how recurrent dynamics are organized into active and weakly evolving regions.
The remainder of the paper is organized as follows. Section~2 introduces the Kolmogorov-flow configuration and numerical formulation, together with the low-dimensional  model, and its validation. Section~3 describes the conventional recurrence search, the latent-space multiple-shooting refinement, and the resulting search performance and converged RPOs. Section~4 examines the spatial organization of the recurrent solutions, including their localization, continuation in Reynolds number, persistence under domain truncation, and transfer of the learned dynamics across domain sizes. Finally, Section~5 summarizes the principal findings and discusses their implications for recurrent dynamics in spatially extended turbulence.

\section{Methodology}
\label{sec:methodology}

\subsection{Kolmogorov flow and numerical formulation}
\label{sec:kflow_formulation}

We consider two-dimensional Kolmogorov flow  in a doubly periodic rectangular domain, $ (x,y)\in[0,L_x]\times[0,L_y]$.
The flow is driven by a sinusoidal body force in the $x$-direction that varies periodically in $y$.  The nondimensional incompressible Navier-Stokes equations are
\begin{align}
    \frac{\partial \boldsymbol{u}}{\partial t}
    + \boldsymbol{u}\cdot\nabla\boldsymbol{u}
    &= -\nabla p
    + \frac{1}{Re}\nabla^2\boldsymbol{u}
    + \sin(n y)\,\boldsymbol{e}_x,
    \label{eq:NS}
    \\
    \nabla\cdot\boldsymbol{u} &= 0,
    \label{eq:incompressibility}
\end{align}
where $\boldsymbol{u}=(u,v)$ is the velocity field, $p$ is the pressure,
$Re$ is the Reynolds number, and $\boldsymbol{e}_x$ is the unit vector in the $x$-direction. 

We define the characteristic length scale as
    $\ell_0={L_x}/{2\pi}$.
The corresponding velocity scale is $U_0=\sqrt{\chi\ell_0}$, where $\chi$ denotes the dimensional forcing amplitude per unit mass. The Reynolds number is therefore defined as
\begin{equation}
    Re
    =\frac{\sqrt{\chi}}{\nu}
     \left(\frac{L_x}{2\pi}\right)^{3/2}.
\end{equation}
The governing equations are solved in the doubly periodic domain
    $(x,y)\in[0,2\pi]\times[0,2\pi/\alpha]$,
with $\alpha={L_x}/{L_y}$.
For the present calculations, $\alpha=1/6$, corresponding to the
nondimensional domain $[0,2\pi]\times[0,12\pi]$.

For computational efficiency, the equations are advanced in vorticity form.
Defining
\[
\omega=(\nabla\times\boldsymbol{u})\cdot\boldsymbol{e}_z
      =\partial_x v-\partial_y u,
\]
the governing equation becomes
\begin{equation}
\frac{\partial \omega}{\partial t}
+\boldsymbol{u}\cdot\nabla\omega
=
\frac{1}{Re}\nabla^2\omega
-n\cos(ny).
\end{equation}

We characterize the flow using the kinetic energy, viscous
dissipation and energy input,
\begin{align}
\mathcal{E}(t)
&=
\frac{1}{2}
\left\langle |\boldsymbol{u}|^2\right\rangle,
\qquad
\mathcal{D}(t)=
\frac{1}{Re}
\left\langle |\nabla\boldsymbol{u}|^2\right\rangle,
\qquad
\mathcal{I}(t)=
\left\langle u\sin(ny)\right\rangle,
\end{align}
where
\begin{equation}
\left\langle q\right\rangle
=
\frac{1}{L_xL_y}
\int_0^{L_x}\int_0^{L_y}
q(x,y)\,{\rm d}y\,{\rm d}x.
\end{equation}

For the laminar solution
\begin{equation}
    \boldsymbol{u}_{\rm lam}
    =
    \frac{Re}{n^2}\sin(ny)\,\boldsymbol{e}_x,
\end{equation}
these quantities take the values
\begin{equation}
    \mathcal{E}_{\rm lam}=\frac{Re^2}{4n^4},
    \qquad
    \mathcal{D}_{\rm lam}=\mathcal{I}_{\rm lam}
    =\frac{Re}{2n^2}.
\end{equation}

The governing equations are solved using the Dedalus
pseudospectral framework \citep{dedalus}. Fourier   discretizations are employed in both periodic directions,  with    $32$ Fourier modes per $2\pi$ in each direction. Thus, the reference $2\pi\times12\pi$ with $n=2$ and $Re=10$ calculation uses $32\times192$
Fourier modes. Comparable spectral resolutions have been used in previous studies of two-dimensional Kolmogorov flow \citep{carlos,ecs_koopman,disdmand}.

To assess grid independence, selected converged RPOs 
were interpolated onto the finer discretization with  64 Fourier modes per $2\pi$, and subsequently reconverged using full-space Newton solver; the converged periods and streamwise shifts were then compared between the two resolutions. For   $RPO_{4.36}$, the relative changes in $T$ and $\ell_x$ are approximately $0.0009\%$ and $0.0044\%$, respectively, indicating that these parameters are insensitive to refinement.

\subsection{Distributed data-driven manifold dynamics}
\label{sec:disdmand}

\begin{figure}
\centering
\includegraphics[width=\linewidth]{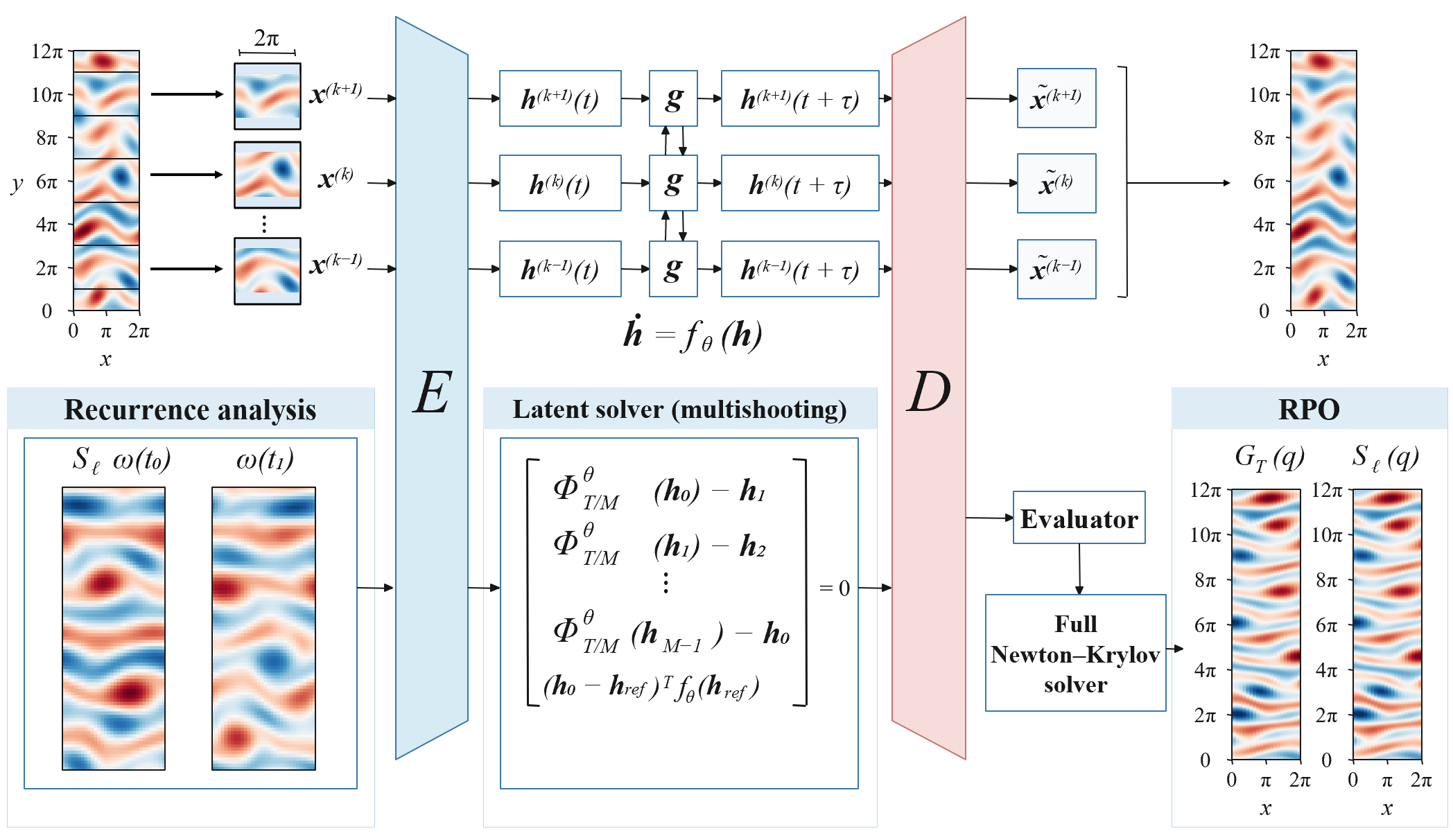}
\caption{
Overview of the framework for identifying RPOs in large-domain Kolmogorov flow.
The upper panels show the construction of DisDManD. DNS snapshots are divided into overlapping spatial patches and mapped by a shared encoder $\mathrm{E}$ to local latent representations. A NODE evolves the latent state of each patch while incorporating interactions with neighboring patches, after which the shared decoder $\mathrm{D}$ reconstructs the physical fields.
The lower panels show the RPO-search procedure. Near-recurrent states identified from DNS are encoded into the latent space and refined using a latent-space multiple-shooting solver. The resulting candidates are decoded to physical space, evaluated using the full governing dynamics, and used as initial guesses for a full Newton-Krylov solver to obtain converged RPOs.
}
\label{fig:framework}
\end{figure}

In this work, we consider systems that are characterized by deterministic, Markovian dynamics, so if $\boldsymbol{x} \in \mathbb{R}^{d_N}$
represents the full state, then the dynamics can be represented by an ODE as
\begin{equation} \label{eq:ODE}
    \dfrac{d\boldsymbol{x}}{dt}=\boldsymbol{f}(\boldsymbol{x}).
\end{equation} 
Here, $\boldsymbol{x}$ represents the full state space 
from highly resolved direct numerical simulation (DNS).
We seek a low-dimensional representation of the sampled turbulent attractor 
of embedding global dimension $d_H$ (i.e., $d_H \ll d_N)$.

Next, we outline briefly  the approach referred to as `Distributed Data-driven Manifold Dynamics' (DisDManD) \citep{disdmand}.
Rather than constructing a single low-dimensional representation of the full state space,  we exploit the spatially extended nature of the flow and represent the state using a collection of  equivalent spatial patches.

The full computational domain is partitioned into $K$ patches, each spanning $2\pi\times2\pi$. For  $[L_x,L_y]=[2\pi,12\pi]$, this gives $K=6$ patches arranged along the $y$-direction.  Since the patch width $2\pi$
is an integer multiple of the forcing period $2\pi/n=\pi$, translations between corresponding patches preserve the phase of the imposed forcing. The patches are therefore related by an exact discrete symmetry of the governing equations. This allows the same local coordinate transformation and local evolution law to be shared across all patches, rather than learning a separate model for each cross-stream location.

For each patch $k$, the encoder maps the local vorticity field $\omega^{(k)}$ to a low-dimensional latent representation $\boldsymbol{h}^{(k)}\in\mathbb{R}^{d_h}$, 
\begin{equation}
    \boldsymbol{h}^{(k)}
    = \mathrm{E}\!\left(\omega^{(k)};\boldsymbol{\theta}_{E}\right),
\end{equation}
while the decoder reconstructs the corresponding physical field according to
\begin{equation}
    \widetilde{\omega}^{(k)}
    = \mathrm{D}\!\left(\boldsymbol{h}^{(k)};
    \boldsymbol{\theta}_{D}\right).
\end{equation}
Here, $\mathrm{E}$ and $\mathrm{D}$ denote the encoder and decoder,
respectively.
The latent variables from all $K$ patches are then concatenated to form the global latent state
\begin{equation}
\boldsymbol{h}
= \left[
\boldsymbol{h}^{(1)},\ldots,\boldsymbol{h}^{(K)}
\right]
\in\mathbb{R}^{d_hK}.
\end{equation}

Because symmetry-equivalent patches share the same encoder and decoder, the network is trained using samples drawn from all spatial patches rather than learning a location-specific representation. This parameter sharing both reduces the number of trainable parameters and exposes the model to a larger ensemble of dynamically equivalent local states. The autoencoder parameters
are obtained by minimizing the mean reconstruction error,

\begin{equation}
\mathcal{L}_{\rm AE}
=
\frac{1}{NK}
\sum_{i=1}^{N}
\sum_{k=1}^{K}
\left\|
\omega^{(k)}(t_i)
-
\mathrm{D}
\left[
\mathrm{E}\left(\omega^{(k)}(t_i)\right)
\right]
\right\|_2^2 .
\label{eq:AE_loss}
\end{equation}

To mitigate sharp changes at the domain boundaries, we overlap the patches by some length $N_{ov}$, effectively expanding the region of each patch.  We then perform a weighted averaging of neighboring patches at the boundary, resulting in continuity of the full state.

After obtaining the local latent representation, the temporal dynamics are modelled directly in latent space.
The essential assumption is that the time evolution of a patch is controlled predominantly by its own state and those of its immediate neighbours.
For patch $k$, the evolution is written in the form
\begin{equation}
\frac{\mathrm{d}\boldsymbol{h}^{(k)}}{\mathrm{d}t}
=
\boldsymbol{g}
\left(
\boldsymbol{h}^{(k-1)},
\boldsymbol{h}^{(k)},
\boldsymbol{h}^{(k+1)};
\boldsymbol{\theta}_{g}
\right),
\qquad k=1,\ldots,K.
\label{NODE_eq}
\end{equation}

The nonlinear mapping $\boldsymbol{g}$ is represented by a neural network. Specifically,  we use a neural ordinary differential equation (NODE) \citep{Chen2019,alec_chaos}. 
During training, all patch states are evolved simultaneously by integrating equation \ref{NODE_eq}. Starting from an encoded DNS state at time $t_i$, the coupled  NODE is integrated over a window containing $N_t$ DNS snapshots, and its trajectory is compared with the corresponding encoded DNS trajectory. The loss may be written schematically as

\begin{equation}
\mathcal{L}_{\rm NODE}
=
\frac{1}{KN_t}
\sum_{k=1}^{K}
\sum_{j=1}^{N_t}
\left\|
\boldsymbol{h}^{(k)}(t_i+j\tau)
-
\widehat{\boldsymbol{h}}^{(k)}(t_i+j \tau)
\right\|_2^2,
\label{eq:NODE_loss}
\end{equation}
where
$\widehat{\boldsymbol{h}}^{(k)}$ denotes the NODE prediction.

The network architectures, training dataset and optimization procedure are given in Appendix B.

\subsubsection{Validation of the learned low-dimensional model for RPO searches}
\label{sec:model_validation}

Before using the learned model to identify recurrent trajectories for $[L_x,L_y]=[2\pi,12\pi]$, we assess its ability to represent states on the turbulent attractor and reproduce the relevant temporal dynamics. To establish a physically meaningful dynamical timescale for assessing the learned dynamics, we estimate the leading Lyapunov exponent from DNS as $\lambda_{\mathrm{LE}}=0.201\pm0.043$, corresponding to a Lyapunov time
$\tau_{LE}=\lambda_{\mathrm{LE}}^{-1}\approx4.99$.
Because chaotic trajectories diverge exponentially, accurate pointwise prediction is not expected to persist far beyond $O(\tau_{LE})$ \citep{chaos_book}.

\begin{figure}
\centering
\includegraphics[width=0.95\textwidth]{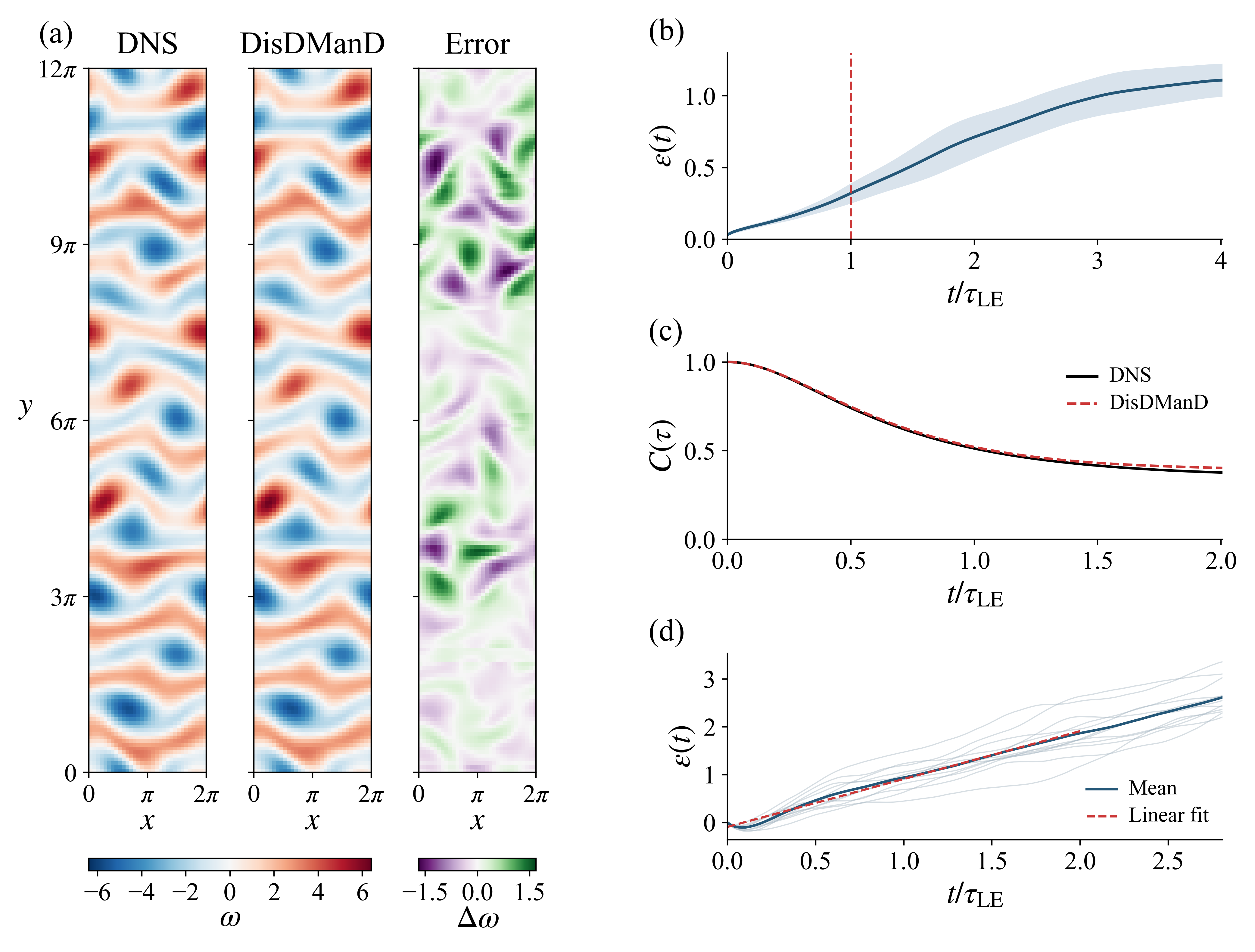}
\caption{
(a) Vorticity fields from DNS and DisDManD at
$t=\tau_{\mathrm{LE}}$.
(b)  Short-time tracking for the DNS and DisDManD averaged over 100 trajectories. The shaded region
indicates trajectory-to-trajectory variability, and the vertical
dashed line marks $t=\tau_{\mathrm{LE}}$.
(c) Temporal autocorrelation $C(t)$ obtained from DNS and DisDManD.
(d) Estimation of the leading Lyapunov exponent from perturbation
growth in the low-dimensional model.
}
\label{fig:DisDManD_validation}
\end{figure}

For $[L_x,L_y]=[2\pi,12\pi]$, the DisDManD model partitions the domain into six $2\pi\times2\pi$ patches, each represented by $d_h=20$ latent variables, yielding a global latent state $\boldsymbol{h}\in\mathbb{R}^{120}$ and reducing the discretized vorticity field from $\mathbb{R}^{6144}$ by a factor of $51.2$. 
Figure~\ref{fig:DisDManD_validation}a compares the DNS vorticity field with the DisDManD prediction after one Lyapunov time, $t=\tau_{LE}$. Despite the substantial dimensionality reduction, the model reproduces the dominant spatial organization of the flow,
while the prediction error remains smaller in amplitude than the underlying vorticity field and is concentrated primarily around regions of strong spatial variation. The error field remains spatially smooth across the boundaries between neighboring patches, with small  discontinuities introduced by the distributed representation.

While figure~\ref{fig:DisDManD_validation}a illustrates the vorticity field prediction for a representative initial condition, figure~\ref{fig:DisDManD_validation}b quantifies the trajectory-tracking error over an ensemble of 100 randomly selected initial conditions. For each trajectory, we define the relative $L_2$ error as 
$    \epsilon(t)
    =
   \left\|
    \omega_{\mathrm{}}(t)
    -
    \widetilde{\omega}_{\mathrm{}}(t)
    \right\|_2/ \left\|
    \omega_{\mathrm{}}(t)
    \right\|_2   $.
The small initial error $\epsilon(t_0)\approx 0.02$, prior to appreciable evolution by the latent dynamics, reflects the accuracy with which the encoder-decoder represents states on the DNS attractor.  The subsequent increase therefore predominantly measures the accumulation of error during the learned time evolution. At $t\simeq\tau_{\mathrm{LE}}$, the prediction error remains substantially below its long-time value, consistent with the strong field-wise correlation observed in figure~\ref{fig:DisDManD_validation}a.  At longer times ($t\approx 3\tau_{LE}$), the error approaches an $\mathcal{O}(1)$ value as initially nearby trajectories decorrelate, as expected for chaotic dynamics.  
Figure~\ref{fig:DisDManD_validation}c compares the temporal autocorrelation $C(\tau)$ obtained from DNS and DisDManD. We define
$C(\tau)
=
{
\left\langle
\left\langle
\omega'(\boldsymbol{x},t)\,
\omega'(\boldsymbol{x},t+\tau)
\right\rangle_{\boldsymbol{x}}
\right\rangle_t
}/{
\left\langle
\left\langle
\omega'(\boldsymbol{x},t)^2
\right\rangle_{\boldsymbol{x}}
\right\rangle_t
},
$
\label{eq:autocorrelation}
where
$\omega'(\boldsymbol{x},t)
=
\omega(\boldsymbol{x},t)
-
\left\langle
\omega(\boldsymbol{x},t)
\right\rangle_t,
$
and $\langle\cdot\rangle_{\boldsymbol{x}}$ and
$\langle\cdot\rangle_t$ denote spatial and temporal averages,
respectively. 
The DNS and DisDManD curves remain in close agreement over the reported interval, indicating that the reduced-order model reproduces the characteristic temporal decorrelation of the full system.

A complementary check for the model is to 
reproduce the intrinsic rate at which nearby trajectories diverge. Figure~\ref{fig:DisDManD_validation}d shows the growth of infinitesimal perturbations in the latent NODE dynamics. A linear fit over the exponential-growth regime gives
$\lambda^{\mathrm{DisDManD}}_{LE}\approx0.201$, corresponding to
$\tau_{LE}^{\mathrm{DisDManD}}\approx4.99$, in close agreement with the DNS estimate
$\lambda^{\mathrm{DNS}}_{LE}=0.201\pm0.043$.
Thus, the low-dimensional model is capable of reproducing the characteristic rate at which chaotic trajectories lose predictability. This agreement suggests that the learned dynamics capture a fundamental instability timescale of the turbulent attractor that is relevant to the subsequent search for recurrent trajectories.

\section{Finding relative periodic orbits}

\subsection{Conventional recurrence searches}
\label{sec:recurrence}

A conventional route to recurrent solutions is to identify pairs of states along a turbulent trajectory that are close in state space and use one state, together with their temporal separation, as an initial guess for a Newton-Krylov solver. We first examine whether this approach remains effective in the spatially extended domain considered here.
The full-state Newton-Krylov procedure follows that used previously by \citet{isaac}. The details relevant to the present RPO calculations are summarized in Appendix~A.

For an RPO, the state recurs after a period $T$ up to a streamwise
translation $\ell_x$,
\begin{equation}
    G_T(\boldsymbol{\omega}_0)
    =
    S_{\ell_x}(\boldsymbol{\omega}_0),
    \label{eq:rpo_condition}
\end{equation}
where $G_T$ denotes the time-$T$ Navier-Stokes flow map and the
translation operator is defined by
\begin{equation}
    S_{\ell_x}\omega(x,y)
    =
    \omega(x-\ell_x,y).
\end{equation}
Equivalently,
\begin{equation}
    \omega(x+\ell_x,y,t+T)
    =
    \omega(x,y,t).
\end{equation}

We therefore quantify the proximity of two turbulent states separated by
$T$ using the normalized recurrence residual

\begin{equation}
R(t,T)
=
\min_{0\leq \ell_x<L_x}
\frac{
\left\|
\omega(x+\ell_x,y,t)-\omega(x,y,t-T)
\right\|_2^2
}{
\left\|\omega(x,y,t)\right\|_2^2
}.
\label{eq:recurrence_physical}
\end{equation}

\begin{figure}
\centering
\includegraphics[width=\linewidth]{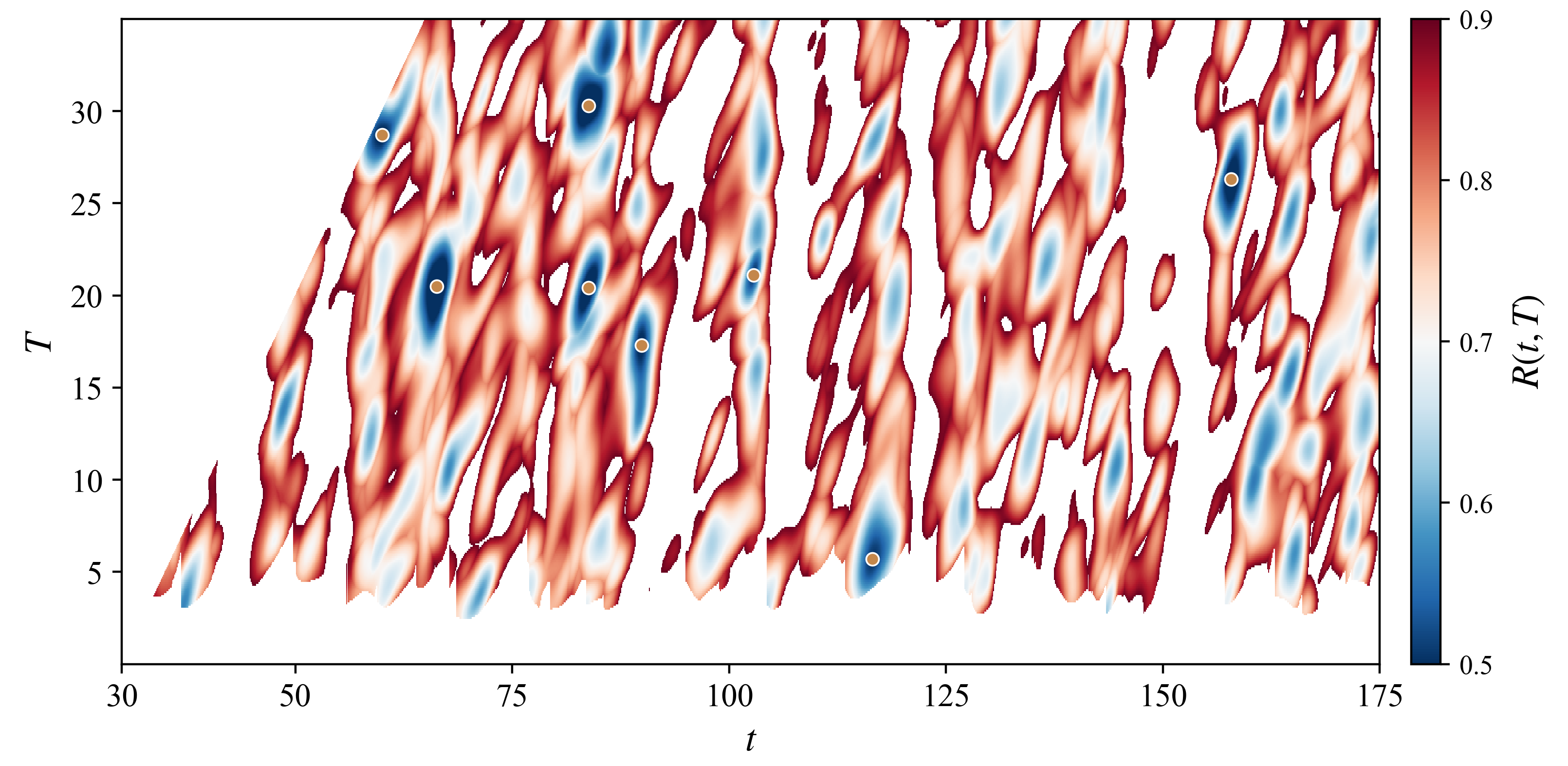}\\
\caption{Recurrence analysis
over $t\in[30,175]$. 
The marked points denote the selected candidate events satisfying the recurrence threshold.}
\label{fig:recurrence_seed001}
\end{figure}

In Fourier space, equation~\eqref{eq:recurrence_physical} becomes
\begin{equation}
R(t,T)
=
\min_{0\leq \ell_x<L_x}
\frac{
\displaystyle\sum_{j,l}
\left|
\Omega_{jl}(t)e^{\mathrm{i}k_{x,j}\ell_x}
-\Omega_{jl}(t-T)
\right|^2
}{
\displaystyle\sum_{j,l}
\left|\Omega_{jl}(t)\right|^2
},
\label{eq:recurrence_fourier}
\end{equation}
where $\Omega_{jl}$ is the Fourier coefficient associated with the streamwise and cross-stream wavenumbers $k_{x,j}$ and $k_{y,l}$, respectively. The minimization accounts for the continuous streamwise translational symmetry. We do not additionally minimize over the discrete cross-stream translations admitted by the Kolmogorov forcing, as for example \citet{chandler_kerswell_2013}.

As an illustration of the recurrence-based candidate selection, figure~\ref{fig:recurrence_seed001} shows the recurrence map
$R(t,T)$ 
over $t\in[30,175]$. Low values of $R(t,T)$ indicate that states separated by a time interval $T$ are close after
optimization over the continuous streamwise shift. 
Candidate near-recurrences, indicated by the markers in figure~\ref{fig:recurrence_seed001}, are identified as local minima satisfying
$R(t,T)<R_{\mathrm{thres}}$, with $R_{\mathrm{thres}}=0.55$.
Each accepted minimum provides an initial guess
$\{\boldsymbol{\omega}(t-T),T,\ell_x\}$ for subsequent refinement. Notably, the selected candidates span recurrence periods comparable to and exceeding the Lyapunov time, where direct shooting becomes increasingly challenging.

\subsection{Latent-space multiple shooting and candidate refinement}
\label{sec:latent_multishooting}

The recurrence analysis yields candidate periods $5\lesssim T\lesssim20$, many exceeding the Lyapunov time. Single shooting would therefore require accurate latent integration over several Lyapunov times, increasing error accumulation and degrading the initial guess supplied to the full-state solver.
We acknowledge that latent single shooting has previously proved effective for recurrent-flow analysis in minimal-flow-unit configurations, which typically exhibit longer Lyapunov times than the spatially extended system considered here (i.e., $\tau_{LE}\approx48$ for plane Couette flow \citep{alec_coutte}, and $\tau_{LE}\approx30.43$ for pipe flow \citep{pipe_flow}).

We therefore adopt a multiple-shooting formulation, which reduces the integration horizon of each shooting segment and limits the accumulation of these errors during the latent refinement. A candidate orbit of period $T$ is divided into $M$ intervals of equal duration
\begin{equation}
 \Delta T = \frac{T}{M},
\end{equation}
where $M$ is chosen such that $\Delta T$ does not substantially exceed the validated NODE prediction horizon. Let
\begin{equation}
\boldsymbol H
=
\left(
\boldsymbol h_0,\boldsymbol h_1,\ldots,
\boldsymbol h_{M-1}
\right),
\end{equation}
where $\boldsymbol h_j$ denotes the latent state at the beginning of the $j$th shooting segment. The states $\boldsymbol h_j$ are treated as independent optimization variables and are initialized by encoding the DNS states at the corresponding segment boundaries of the candidate recurrence. Let
$\Phi_\tau^\theta$ denote the flow map generated by the trained NODE,
defined by
\begin{equation}
    \frac{\mathrm d\boldsymbol h}{\mathrm dt}
    =
    \boldsymbol f_\theta(\boldsymbol h).
\end{equation}
The complete vector of unknowns is
\begin{equation}
    \boldsymbol q
    =
    \left(
    \boldsymbol h_0,\ldots,\boldsymbol h_{M-1},T
    \right).
\end{equation}
The latent multiple-shooting equations are then
\begin{equation}
\boldsymbol F_{\mathrm{MS}}(\boldsymbol q)
=
\begin{bmatrix}
\Phi_{\Delta T}^{\theta}(\boldsymbol h_0)-\boldsymbol h_1
\\[1mm]
\Phi_{\Delta T}^{\theta}(\boldsymbol h_1)-\boldsymbol h_2
\\
\vdots
\\
\Phi_{\Delta T}^{\theta}(\boldsymbol h_{M-2})
-\boldsymbol h_{M-1}
\\[1mm]
\Phi_{\Delta T}^{\theta}(\boldsymbol h_{M-1})
-\boldsymbol h_0
\\[1mm]
\psi_t(\boldsymbol h_0)
\end{bmatrix}
=
\boldsymbol 0,
\label{eq:latent_multishooting}
\end{equation}
where the first $M-1$ vector equations enforce continuity between successive segments and the final vector equation imposes periodic closure. Because the origin of time along a periodic orbit is arbitrary, we impose the temporal phase condition
\begin{equation}
    \psi_t(\boldsymbol h_0)
    =
    \left(
    \boldsymbol h_0-\boldsymbol h_{\mathrm{ref}}
    \right)^{\mathsf T}
    \boldsymbol f_\theta(\boldsymbol h_{\mathrm{ref}})
    =
    0,
\label{eq:temporal_phase}
\end{equation}
where $\boldsymbol h_{\mathrm{ref}}$ is a fixed reference state taken from the candidate trajectory. Then, 
equation~\eqref{eq:latent_multishooting} supplies $MKd_h+1$ equations for the $MKd_h+1$ unknown components of $\boldsymbol q$.

The present latent formulation imposes strict periodic closure and therefore searches for periodic orbits rather than relative periodic orbits. In particular, streamwise translations are not included as continuous optimization variables in the latent-space problem.
This restriction is a limitation of the present reduced-order search and prevents exact latent closure of recurrences whose periodicity requires a non-zero
spatial shift. Latent optimisation occasionally converged toward the trivial limit $T\rightarrow0$. We therefore discarded candidates whose optimised period fell below $T=0.1$.

The convergence of the latent multiple-shooting iteration is monitored using
the normalized residual
\begin{equation}
r_{\mathrm{MS}}(\boldsymbol q)
=
\left[
\frac{
\displaystyle
\sum_{i=0}^{M-1}
\left\|
\Phi_{\Delta T}^{\theta}(\boldsymbol h_i)
-\boldsymbol h_{i+1}
\right\|_2^2
}{
\displaystyle
\sum_{i=0}^{M-1}
\left\|\boldsymbol h_i\right\|_2^2
}
\right]^{1/2},
\qquad
\boldsymbol h_M\equiv\boldsymbol h_0.
\label{eq:latent_ms_residual}
\end{equation}
Thus, $r_{\mathrm{MS}}=0$ corresponds to exact continuity between all successive shooting segments, including closure of the final segment onto
the initial latent state.

Each near-recurrence identified in \S\ref{sec:recurrence} provides an initial physical state $\omega_0^{(0)}$ and an estimated recurrence period $T^{(0)}$. The state is first mapped to the latent representation,
$\boldsymbol h_0^{(0)}=\mathrm{E}(\omega_0^{(0)})$, which is then refined together with the recurrence time through latent optimization. This yields an optimized latent initial state $\boldsymbol h_0^*$ and interval duration $\Delta T^*$, corresponding to a total period $T^*=M\Delta T^*$. The refined latent state is subsequently decoded to obtain the physical-space initial condition $\omega_0^*=\mathrm{D}(\boldsymbol h_0^*)$, which generally differs from the original DNS state $\omega_0^{(0)}$. Because both the initial state and period are modified during this refinement, the streamwise shift that best satisfies the recurrence under the full dynamics may also differ from that associated with the original candidate.



\subsubsection{Physical-space evaluation of optimized candidates}
\label{sec:physical_evaluation}

A small latent-space residual does not necessarily imply that the decoded
state is recurrent under the governing Navier-Stokes  equations.
Each optimized
candidate is therefore evaluated using the full physical flow map $G_T$. Starting from  the decoded state $\boldsymbol{\omega}_0^*$, the Navier-Stokes equations are integrated over the optimized period $T^*$. The corresponding streamwise shift is determined by minimizing the mismatch between the time-evolved field and the translated initial state,
\begin{equation}
\ell_x^*
=
\underset{-L_x/2\leq\ell_x<L_x/2}{\operatorname{arg\,min}}\,
\left\|
G_{T^*}(\boldsymbol{\omega}_0^*)
-
S_{\ell_x}(\boldsymbol{\omega}_0^*)
\right\|_2^2.
\label{eq:optimized_shift}
\end{equation}
The corresponding physical-space recurrence residual is
\begin{equation}
R_{\mathrm{phys}}^*
=
\frac{
\left\|
G_{T^*}(\boldsymbol{\omega}_0^*)
-
S_{\ell_x^*}(\boldsymbol{\omega}_0^*)
\right\|_2^2
}{
\left\|\boldsymbol{\omega}_0^*\right\|_2^2
}.
\label{eq:decoded_physical_residual}
\end{equation}
Thus, $r_{\mathrm{MS}}$ measures closure under the learned latent
dynamics, whereas $R_{\mathrm{phys}}^*$ measures recurrence of the decoded
candidate under the full Navier-Stokes dynamics.

The optimized candidates are ranked according to
$R_{\mathrm{phys}}^*$, and those with the smallest physical-space residuals are retained for full-state convergence. This evaluation is essential because errors in the learned dynamics and in the encoder-decoder reconstruction can cause a state with a small latent residual to have a substantially larger physical-space residual.

In summary, for each selected candidate, the triplet
$ \left(\boldsymbol{\omega}_0^*,T^*,\ell_x^*\right)$
is supplied as the initial guess to the full-state Newton-Krylov solver.
The learned model is therefore used only to construct and screen initial guesses; all reported recurrent solutions satisfy the governing Navier-Stokes equations to numerical precision.

\subsection{Search performance and converged solutions}

\label{sec:initial_guess_evaluation}

\begin{table}
\centering
\begin{tabular}{lccccc}
\hline
Candidate-selection method
& $M$
& Candidates screened
& Sent to full-state Newton
& RPOs found
& Success rate \\[3pt]
\hline
Full-space recurrence
& - & 10 & 10 & 0 & 0\% \\[3pt]
Latent-assisted
& 1 & 98  & 22  & 5 & 23\% \\
& 2 & 189 & 28 & 7 & 25\% \\
& 3 & 235 & 19  & 2 & 11\% \\[3pt]
Latent-assisted total
& - & 522 & 69 & 14 & 20\% \\
\hline
\end{tabular}
\caption{Comparison of full-space recurrence and latent-assisted candidate
selection. The success rate is the fraction of candidates sent to the
full-state Newton-Krylov solver that converged to an RPO. }
\label{tab:ecs_search_comparison}
\end{table}

\begin{figure}
\centering
\includegraphics[width=\textwidth]{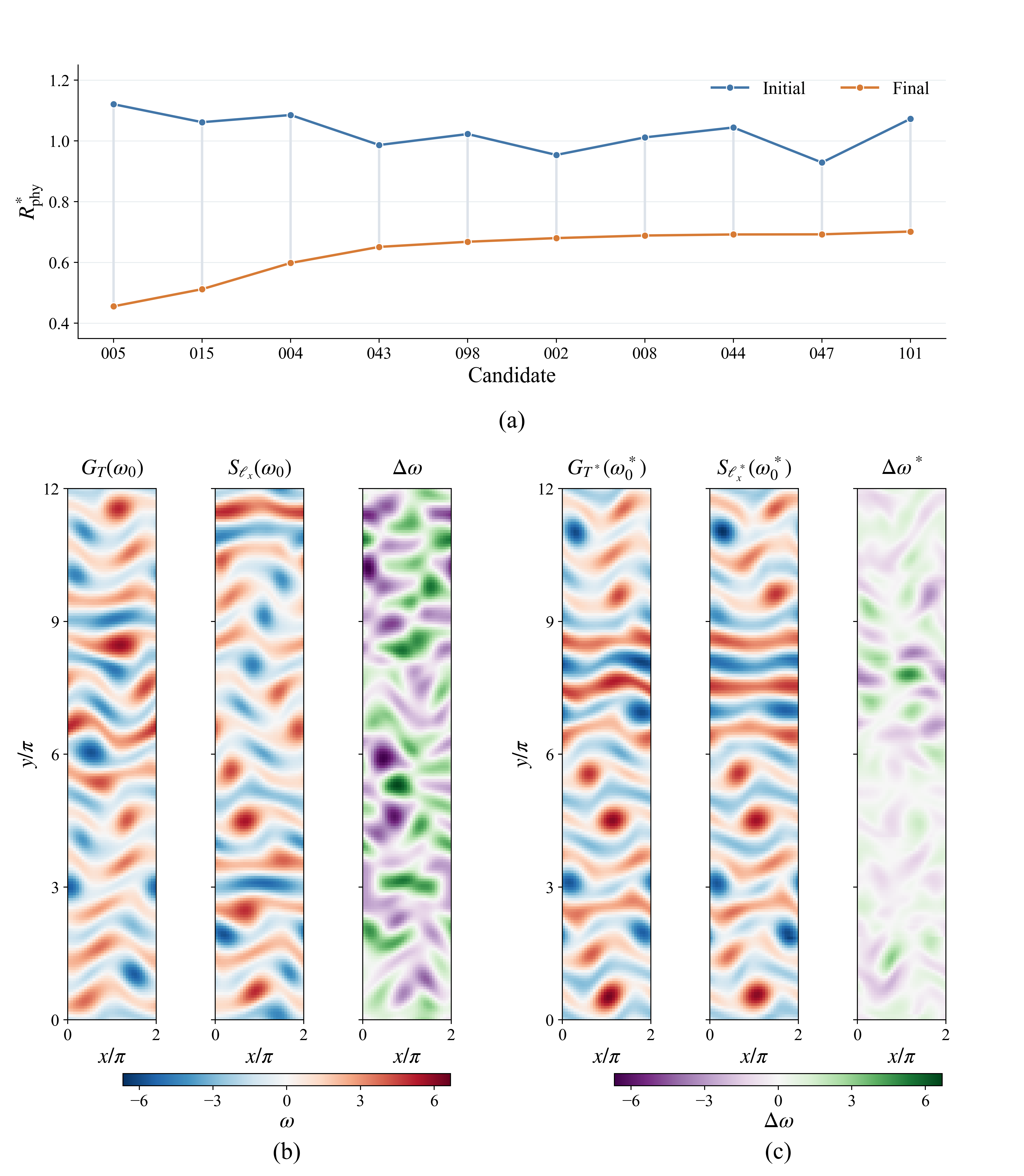}
\caption{
Improvement of recurrence candidates 
through latent-space multishooting.
(a) Physical-space recurrence residual, $R_{\mathrm{phys}}^*$, before and after multishooting optimization for ten selected candidates   with the smallest final residuals,
ranked according to $R_{\mathrm{phys}}^*$.
In panel (b), the evolved state $G_T(\omega_0)$ is compared with the shifted initial state $S_{\ell_x}(\omega_0)$, whereas panel (c) shows
the corresponding optimized states $G_{T^*}(\omega_0^*)$ and
$S_{\ell_x^*}(\omega_0^*)$.
This case corresponds to candidate 005 from panel (a).
}
\label{fig:latent_refinement_before_after}
\end{figure}

Table~\ref{tab:ecs_search_comparison} summarises the global and latent-assisted search statistics.
None of the 10 candidates selected directly from the full-space recurrence analysis converged to an ECS.
We acknowledge that the number of candidates is smaller than the data-driven case.
However, the ten candidates considered here were already
the most promising ones identified by the recurrence analysis according to the
criterion described in Section.~\ref{sec:recurrence}. Applying the full-state Newton-Krylov solver
indiscriminately to the much larger recurrence candidate set would be
computationally prohibitive (as discussed below), particularly because candidate periods frequently
extend to $T\approx 20$-$30$, as shown in figure~\ref{fig:recurrence_seed001}.

By contrast, the latent-space search screened 98, 189, and 235 recurrence candidates using $M=1$, $2$, and $3$ shooting intervals, respectively. Only the most promising 22, 28, and 19 candidates were passed to the substantially more expensive
full-state Newton-Krylov solver. Of these 69 candidates, 14 converged to relative periodic orbits, giving an overall success rate of $20\%$.

The computational advantage can be quantified from the initial search campaign, which yielded the first ten RPOs reported here. Consider, in particular, the $M=2$ search.
Latent optimization of 102 candidates required approximately $7$~h in total, after which only 19 candidates were passed to the full-state Newton-Krylov solver. A single full-state calculation typically required $4$-$7$~h/case; thus
applying Newton-Krylov directly  to all 102 candidates would  have required approximately $408$-$714$~h. 
The latent-assisted search instead required approximately $83$-$140$~h, including the latent optimization, corresponding to a reduction in computational cost by approximately a factor of five. Of the 19 candidates passed to the full-state solver, five converged to distinct RPOs.
All computations were performed on the NCSA Delta cluster using AMD EPYC 7763 (Milan) CPUs. The 
calculations used 12 allocated CPU cores.

Latent-space multiple shooting thus serves as both a refinement procedure and an efficient filter for full-state calculations. The $M=2$ search gave the highest yield and conversion rate, with seven RPOs from 28 Newton-Krylov calculations. Although $M=3$ was less efficient overall, the additional shooting intervals shorten the latent integration horizon and are therefore advantageous for long-period recurrences. As shown below, the resulting catalogue includes RPOs with $T>2\tau_{LE}$.

Figure~\ref{fig:latent_refinement_before_after}a compares the
physical-space residual before and after latent-space optimization for the fourteen candidates with the smallest values of $R_{\mathrm{phys}}^*$, ordered by their final residual. The latent refinement reduces the physical-space error for all fourteen candidates, in some cases by more than a factor of two. Nevertheless, the refined residuals remain appreciable,
emphasizing that the latent optimization is used to identify and improve promising initial guesses rather than to produce converged RPOs directly.

Figures~\ref{fig:latent_refinement_before_after}b-c show the physical-space fields for   candidate 005 from panel~\ref{fig:latent_refinement_before_after}a. Before optimization, the Navier-Stokes-evolved field $G_{T^{(0)}}(\boldsymbol{\omega}_0^{(0)})$ differs substantially from the optimally shifted initial field $S_{\ell_x^{(0)}}(\boldsymbol{\omega}_0^{(0)})$ (see figure \ref{fig:latent_refinement_before_after}b). After latent optimization, the decoded state $\boldsymbol{\omega}_0^*=\mathcal{D}(\boldsymbol{h}_0^*)$, together with the
optimized period $T^*$ and shift $\ell_x^*$, produces much closer agreement between $G_{T^*}(\boldsymbol{\omega}_0^*)$ and $S_{\ell_x^*}(\boldsymbol{\omega}_0^*)$ (see figure \ref{fig:latent_refinement_before_after}c). The associated difference fields,
\begin{equation}
\Delta\omega^{(0)}
= G_{T^{(0)}}(\boldsymbol{\omega}_0^{(0)})
- S_{\ell_x^{(0)}}(\boldsymbol{\omega}_0^{(0)}),
\end{equation}

and

\begin{equation}
\Delta\omega^* = G_{T^*}(\boldsymbol{\omega}_0^*) - S_{\ell_x^*}(\boldsymbol{\omega}_0^*),
\end{equation}
make the reduction in the spatial mismatch explicit.

Latent multiple shooting therefore does more than rank
the original recurrence candidates; it modifies the initial condition, period, and streamwise shift to produce an improved seed for the full-state solver. Following decoding, the optimal streamwise shift is recomputed from the full Navier-Stokes evolution. The remaining mismatch is subsequently removed by Newton-Krylov refinement, and a candidate is accepted as an RPO only when the full-state RPO residual reaches machine precision tolerance.

Multiple shooting therefore makes it possible to explore recurrence periods that exceed the prediction horizon of any individual NODE trajectory, while retaining a substantially lower computational cost than full-state optimization.

\section{Spatial structure and localization of the relative periodic orbits}
\label{sec:rpo_results}

\begin{table}
\centering
\begin{center}
\renewcommand{\arraystretch}{1.25}
\setlength{\tabcolsep}{7pt}

\begin{tabular}{lcccccc}
RPO & $T$ & $\ell_x$ &  $R$ & $M$ & $\Lambda_A$ & Character \\[5pt]
\hline

$\mathrm{RPO}_{3.65}$  & 3.65  & -0.53  & $3.4\times10^{-12}$ & 2 & 0.21 & localized \\
$\mathrm{RPO}_{4.36}$  & 4.36  & -0.06 & $1.6\times10^{-12}$ & 2 & 0.29 & localized \\
$\mathrm{RPO}_{4.51}$  & 4.51  & 0.50   & $1.3\times10^{-13}$ & 1 & 0.32 & localized \\
$\mathrm{RPO}_{4.67}$  & 4.67  & -1.05    & $4.8\times10^{-12}$ & 1 & 0.48 & localized \\
$\mathrm{RPO}_{6.01}$  & 6.01  & 0.04  & $6.0\times10^{-12}$ & 1 & 0.36 & localized \\

$\mathrm{RPO}_{6.30}$  & 6.30  & 0.36   & $3.1\times10^{-11}$ & 2 & 0.53 & localized \\
$\mathrm{RPO}_{10.75}$ & 10.75 & -0.03 & $2.6\times10^{-11}$ & 2 & 0.90 & fully active \\
$\mathrm{RPO}_{11.78}$ & 11.78 & 0.05  & $5.3\times10^{-12}$ & 2 & 0.99 & fully active \\
$\mathrm{RPO}_{11.87}$ & 11.87 & -0.02 & $2.6\times10^{-13}$ & 1 & 0.97 & fully active \\
$\mathrm{RPO}_{12.71}$ & 12.71 & 2.98   & $1.2\times10^{-13}$ & 2 & 1.00 & fully active \\
$\mathrm{RPO}_{14.02}$ & 14.02 & -0.02 & $1.5\times10^{-12}$ & 1 & 1.00 & fully active \\
$\mathrm{RPO}_{14.10}$ & 14.10 & -0.88  & $3.2\times10^{-14}$ & 3 & 0.93 & fully active \\

$\mathrm{RPO}_{14.81}$ & 14.81 & 0.49   & $1.2\times10^{-11}$ & 3 & 1.00 & fully active \\
$\mathrm{RPO}_{16.45}$ & 16.45 & 1.59   & $2.2\times10^{-11}$ & 3 & 1.00 & fully active \\

\end{tabular}

\caption{ 
Relative periodic orbits, ordered by increasing period $T$.
Here $\ell_x$ is the streamwise shift, $R$ is the tolerance,
$M$ is the number of shooting intervals,  and $\Lambda_A$ denotes the active region fraction. 
}
\label{tab:rpo_solutions}

\end{center}
\end{table}

\subsection{Catalogue of converged solutions}
\label{sec:rpo_catalogue}

The latent-assisted search yielded fourteen distinct RPOs after full-state Newton-Krylov refinement, as summarized in Table~\ref{tab:rpo_solutions}. 
Their periods span $3.65\leq T\leq16.45$, while their streamwise shifts range from $\ell_x=-1.05$ to $2.98$. 
Ten of the fourteen solutions have $T>\tau_{LE}$, and eight have $T>2\tau_{LE}$, demonstrating that the search is not restricted to short recurrences close to the Lyapunov timescale. In particular, the longest orbit has $T=16.45$. The recovery of these
long-period solutions highlights the role of latent-space multiple shooting. 
For reference, \citet{Zhigunov_Page_2026} reported RPOs with periods up to $T=3.91$ for their $Re=40$, $n=4$, and $[L_x,L_y]=[2\pi,4\pi]$ configuration; the present catalogue contains substantially longer recurrence periods, albeit at different flow parameters. As shown below, the shorter-period RPOs tend to be spatially localized, whereas the longer-period solutions involve activity over most of the domain.

\begin{figure}
\centering
\includegraphics[width=\linewidth]{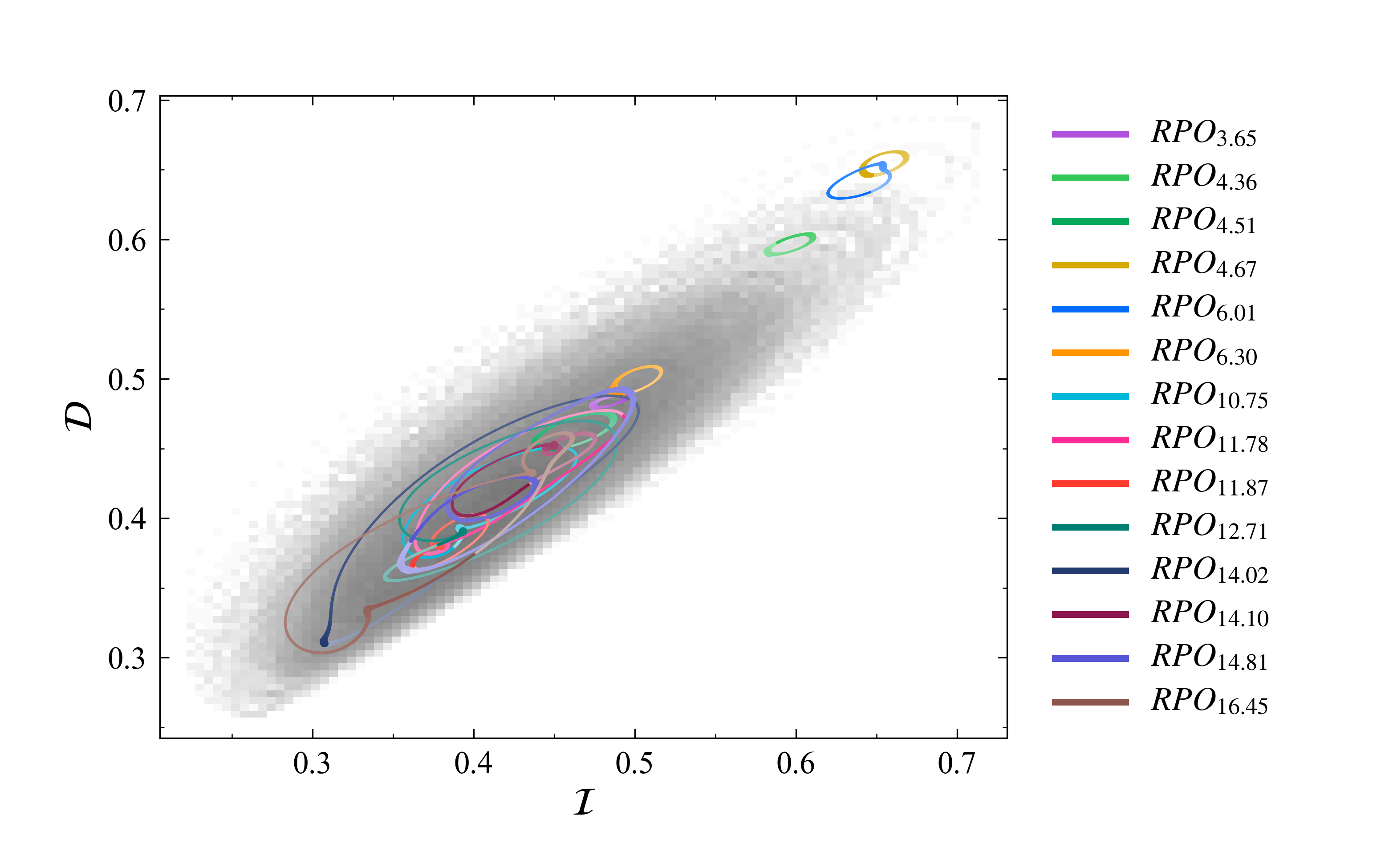}
\caption{
Trajectories of the fourteen converged RPOs projected onto the input-dissipation plane.
The grey background shows the distribution sampled by a long DNS trajectory. 
}
\label{fig:ID_rpo}
\end{figure}

Figure~\ref{fig:ID_rpo} projects the fourteen converged RPOs onto the input-dissipation plane, superimposed on the joint probability density. Most of the RPOs occupy the high-probability core of the turbulent attractor in this projection. A smaller subset extends into the
upper-right tail of the distribution: $\mathrm{RPO}_{4.36}$,
$\mathrm{RPO}_{6.01}$, and $\mathrm{RPO}_{4.67}$ sample progressively larger input and dissipation. Thus, the search recovers recurrent dynamics both within the statistically dominant region and during less frequently visited high-input, high-dissipation excursions.

Rather than pursuing an exhaustive catalogue, we use this set of fourteen RPOs to investigate the physical organization of recurrence in the extended domain. The set includes solutions from both the densely sampled core and the less frequently visited high-input, high-dissipation region of the turbulent distribution, providing a sufficiently varied basis for examining spatial localization in the next sections.

\subsection{Spatial localization and its Re dependence}
\label{sec:localization}

The converged RPOs in table~\ref{tab:rpo_solutions} exhibit a range of spatial organizations, from solutions in which temporal modulation extends throughout most of the domain to others in which it is confined to a restricted region. To quantify this localization, we measure how the local fluctuation intensity varies over one period. At each transverse position $y$, we define
\begin{equation}
A(y,t)
=
\left[
\frac{1}{L_x}
\int_0^{L_x}
\left|
\omega(x,y,t)-\overline{\omega}(x,y)
\right|^2
\,\mathrm{d}x
\right]^{1/2},
\end{equation}
where
\begin{equation}
\overline{\omega}(x,y)
=
\frac{1}{T}\int_0^T\omega(x,y,t)\,\mathrm{d}t
\end{equation}
is the temporal mean of the RPO. Thus, $A(y,t)$ measures the instantaneous amplitude of the local vorticity fluctuations about the orbit mean.
 We then quantify the temporal modulation of this amplitude through
\begin{equation}
    \sigma_A(y)
    =
    \left[
    \frac{1}{T}
    \int_0^T
    \left(A(y,t)-\overline{A}(y)\right)^2
    \,\mathrm{d}t
    \right]^{1/2},
    \qquad
    \overline{A}(y)
    =
    \frac{1}{T}\int_0^T A(y,t)\,\mathrm{d}t .
\end{equation}

We use $\sigma_A(y)$ as a measure of temporal modulation of the local fluctuation amplitude. A small value of $\sigma_A(y)$ does not by itself imply that the local vorticity field is steady; it indicates
only that the magnitude of its fluctuations varies weakly over the recurrent cycle.

\begin{figure}
\centering
\includegraphics[
width=\textwidth]{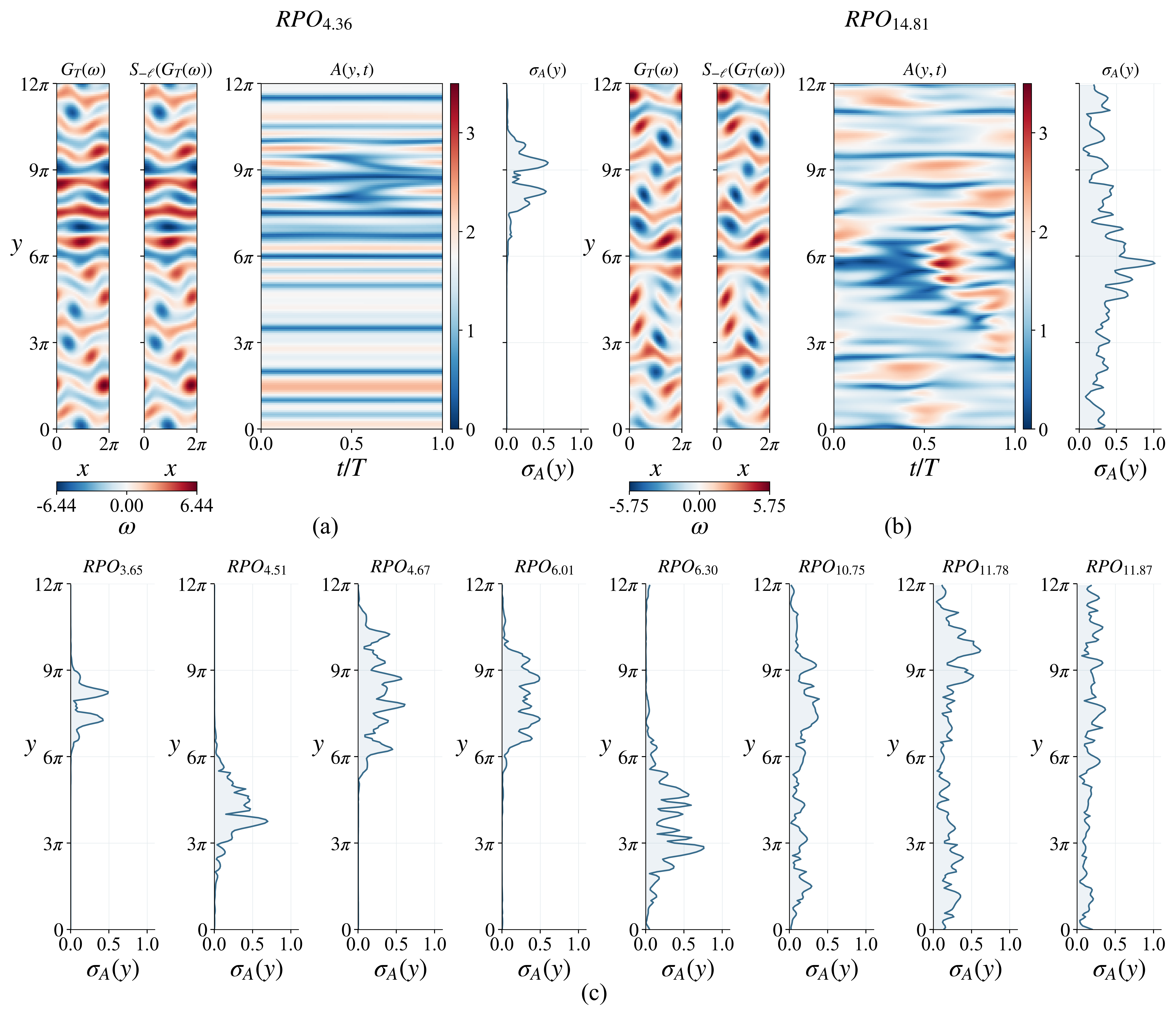}
\caption{
Spatial localization of the converged RPOs. 
(a,b) Two representative cases, $\mathrm{RPO}_{4.36}$ and
$\mathrm{RPO}_{14.81}$, respectively. For each orbit, the first two columns show the evolved vorticity field and the corresponding field after compensation for the streamwise shift. The third column shows the spatio-temporal fluctuation amplitude $A(y,t)$ over one period, and the fourth column shows the corresponding activity profile $\sigma_A(y)$.
(c) Activity profiles $\sigma_A(y)$ for eight additional RPOs.
}
\label{fig:rpo_localization}
\end{figure}

Figure~\ref{fig:rpo_localization} illustrates the contrast using $\mathrm{RPO}_{4.36}$ and $\mathrm{RPO}_{14.81}$ as representative localized and domain-filling cases, respectively. For $\mathrm{RPO}_{4.36}$, strong temporal modulation is concentrated primarily within $8\pi\lesssim y\lesssim10\pi$, producing a pronounced peak in $\sigma_A(y)$. Outside this interval, the fluctuation intensity is only weakly modulated over the orbit (see figure~\ref{fig:rpo_localization}a). By contrast, $\mathrm{RPO}_{14.81}$ exhibits appreciable modulation across most of the cross-stream extent, with no comparably extended region of weak modulation (see figure~\ref{fig:rpo_localization}b). The $\sigma_A(y)$ profiles of other eight  RPOs span a continuous range between these limits (see figure~\ref{fig:rpo_localization}c). Some exhibit one or more confined regions of strong modulation, whereas others show appreciable modulation over most of the domain. Thus, the temporal modulation associated with an RPO can be strongly localized even though the RPO itself is a global solution of the governing equations.

To quantify this localization across the RPO catalogue, we define the active-region indicator
\begin{equation}
\chi_A(y)= \mathcal{H}\!\left[\sigma_A(y)-\sigma_c\right],
\end{equation}
where $\mathcal{H}$ is the Heaviside function. The threshold is
$\sigma_c=0.0433$, applied uniformly to all RPOs, so that regions with $\sigma_A(y)>\sigma_c$ are classified as active and those with $\sigma_A(y)\leq\sigma_c$ as inactive. Here, ``active'' refers specifically to appreciable temporal modulation of the local fluctuation amplitude; an inactive region need not therefore be laminar or strictly steady.

The spatial extent of the active region and the corresponding active fraction are
\begin{equation}
L_A=\int_0^{L_y}\chi_A(y)\,\mathrm{d}y,
\qquad
\Lambda_A=\frac{L_A}{L_y}.
\label{eq:active_extent}
\end{equation}
Accordingly, \(\Lambda_A\) quantifies the cross-stream extent of the temporally active region, with \(\Lambda_A\simeq1\) indicating activity across most of the domain and smaller values indicating stronger localization.

Table \ref{tab:rpo_solutions}
also reports the active fraction $\Lambda_A$ for the converged RPOs, revealing  substantial variation in the spatial extent of the recurrent dynamics.
$\mathrm{RPO}_{3.65}$ and $\mathrm{RPO}_{4.36}$ are the most strongly localized, with $\Lambda_A\approx0.2$ and $0.3$, respectively;
whereas the RPOs with longer periods are characterised 
by $\Lambda_A\approx1$.
Within this set of 14 RPOs, we can observe that  the
longer-period RPOs are more likely to be fully active,  whereas shorter-period RPOs tend to consist of localized active regions patched together with inactive
regions.
This suggests that increasing recurrence time is associated with a broader spatial organization of the temporal modulation.

\begin{figure}
\centering
\includegraphics[width=0.9\linewidth]{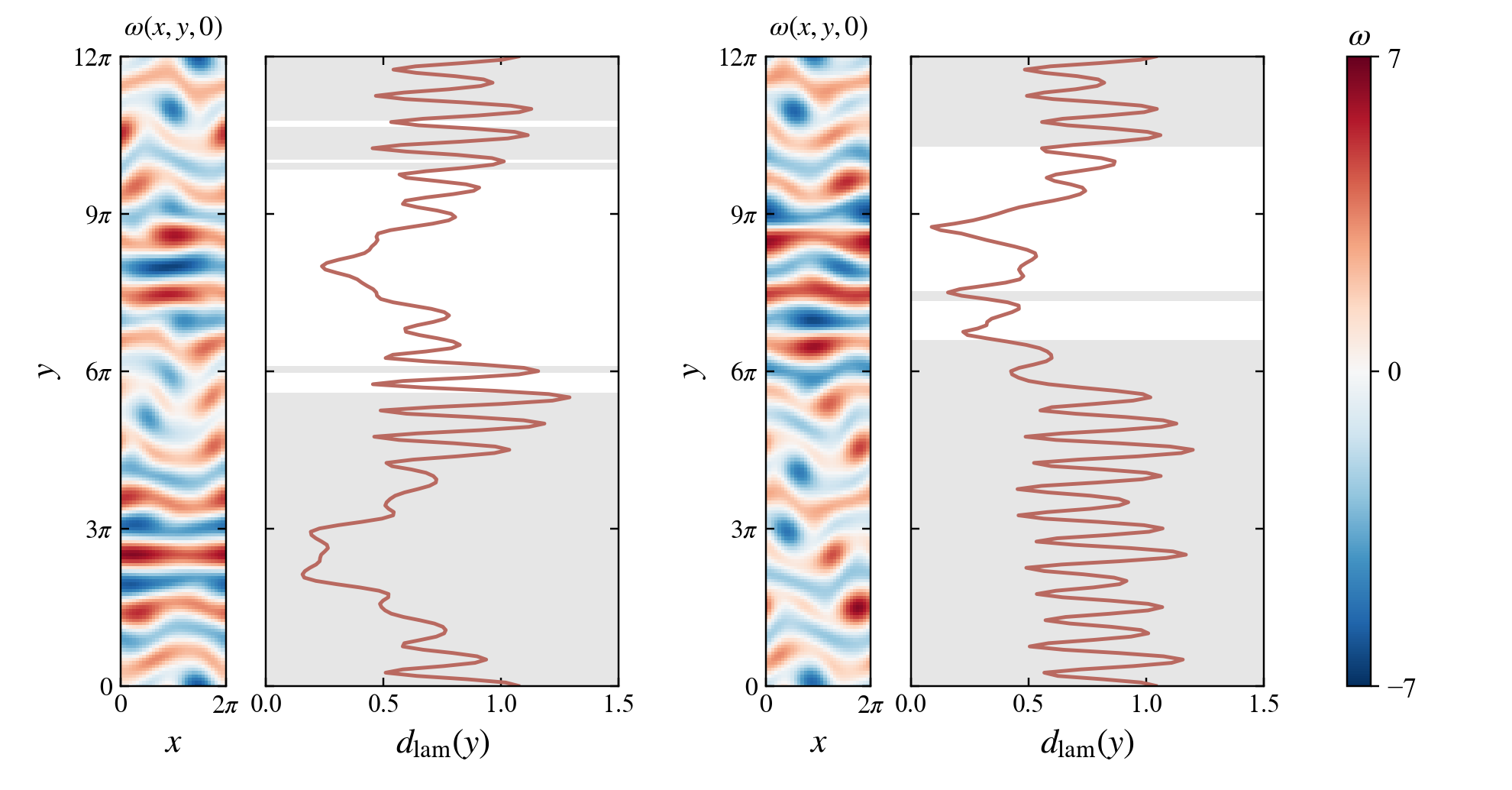}
\caption{
Comparison of the inactive regions with the laminar state for $\mathrm{RPO}_{6.01}$ and $\mathrm{RPO}_{4.36}$. For each RPO, the left panel shows the instantaneous vorticity field $\omega(x,y,0)$, while the adjacent panel shows $d_{\mathrm{lam}}(y)$, computed over one RPO period. The shaded regions denote the inactive regions.
}
\label{fig:RPO_dlam}
\end{figure}

To assess whether the weakly modulated regions approach the laminar state, we define the local distance
\begin{equation}
d_{\rm lam}(y)
=
\left[
\left\langle
|\omega(x,y,t)-\omega_{\rm lam}(x,y)|^2
\right\rangle_{x,t}
\right]^{1/2}.
\label{eq:dlam}
\end{equation}
Here, $d_{\rm lam}(y)$ measures the local root-mean-square difference  from the laminar solution.

Figure \ref{fig:RPO_dlam} shows $d_{\rm lam}(y)$ for two representative RPOs, with the inactive regions identified from $\sigma_A(y)$ shaded. 
For $\mathrm{RPO}_{6.01}$, $d_{\rm lam}(y)$ exhibits a pronounced minimum near $y\simeq3\pi$, where the instantaneous vorticity field  locally approaches the laminar profile. Away from this narrow region, however, the inactive region
remains appreciably separated from the laminar state. This distinction is even clearer for $\mathrm{RPO}_{4.36}$, whose inactive surroundings retain
substantial spatial structure despite their weak temporal modulation. Thus, the inactive regions, in this work, are  regions of weak temporal activity,
rather than  proximity to the laminar solution.
The corresponding profiles for the remaining RPOs are reported in Appendix~C.

\begin{figure}
\centering
\includegraphics[width=\linewidth]{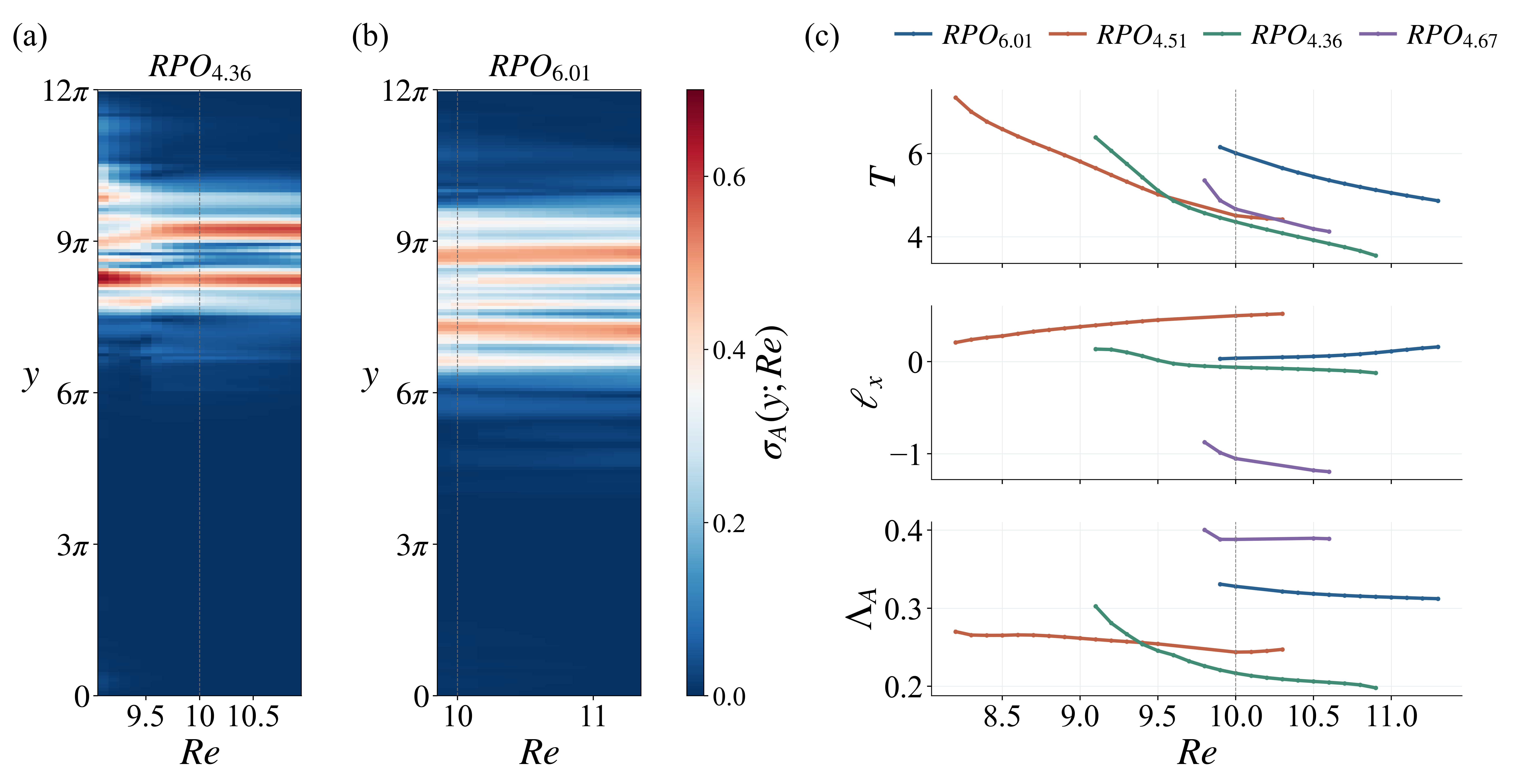}
\caption{
Continuation of selected RPOs with Reynolds number.
(a,b) Evolution of the localization profile $\sigma_A(y;Re)$ along the continuation branches of $\mathrm{RPO}_{4.36}$ and $\mathrm{RPO}_{6.01}$, respectively. The vertical dashed line marks the reference case at $Re=10$.
(c) Variation of the period $T$ (top), streamwise shift $\ell_x$ (middle), and normalized active length $\Lambda_A$ (bottom) with $Re$ for $\mathrm{RPO}_{6.01}$, $\mathrm{RPO}_{4.51}$, $\mathrm{RPO}_{4.36}$, and $\mathrm{RPO}_{4.67}$.
}
\label{fig:Re_continuation}
\end{figure}

To determine whether localization persists away from the reference conditions, $\mathrm{RPO}_{6.01}$, $\mathrm{RPO}_{4.51}$, $\mathrm{RPO}_{4.36}$ and $\mathrm{RPO}_{4.67}$ were continued from
$Re=10$ using natural-parameter continuation. The Reynolds number was varied in increments of $\Delta Re=0.1$ and the step halved if convergence failed, with each converged solution used as the initial guess for the Newton-Krylov solve at the subsequent parameter value. 
The continuation was performed over a limited range of $Re$ both above and below the reference state.

Figures~\ref{fig:Re_continuation}a,b show that the spatial localization of the continued RPOs persists over the accessible range of Reynolds numbers.
For both $\mathrm{RPO}_{4.36}$ and $\mathrm{RPO}_{6.01}$, the region of strongest temporal modulation remains centred at approximately the same cross-stream location as $Re$ is varied. The principal difference is in its extent:  the active band of $\mathrm{RPO}_{4.36}$ narrows appreciably with increasing $Re$, whereas that of $\mathrm{RPO}_{6.01}$ changes comparatively little.
Thus, for these two RPOs, continuation changes the strength and width of the localized dynamics without producing an appreciable displacement of their cross-stream position.

A clearer separation between temporal and spatial responses emerges from figure~\ref{fig:Re_continuation}c. The period decreases monotonically with $Re$ along all four branches, whereas the active fraction $\Lambda_A$ is
considerably less sensitive and varies in a branch-dependent manner. For $\mathrm{RPO}_{6.01}$, $\mathrm{RPO}_{4.51}$ and $\mathrm{RPO}_{4.67}$, $\Lambda_A$ changes only modestly over the computed
continuation interval. By contrast, $\mathrm{RPO}_{4.36}$ decreases from $\Lambda_A\simeq0.30$ to $\simeq0.20$, consistent with the narrowing of the
active band in figure~\ref{fig:Re_continuation}a. Hence, the systematic shortening of the recurrence period is not accompanied by a universal change in spatial extent; the temporal timescale varies similarly across
the branches, while their degree of localization responds differently.

The streamwise shift provides a further independent branch property. Its variation with $Re$ differs among the four solutions and shows no systematic relation to either $T$ or $\Lambda_A$. Along the $\mathrm{RPO}_{4.36}$ branch, the continuously tracked
streamwise shift $\ell_x$ passes through zero, corresponding to a reversal of the mean streamwise drift $c_x=\ell_x/T$, while the cross-stream position of the active region remains approximately unchanged. Thus, changes in streamwise drift are not necessarily accompanied by a displacement of the localized active region.

\subsection{Persistence of active RPO regions under domain truncation}
\label{sec:reduced_domains}

\begin{figure*}
\centering
\includegraphics[width=\textwidth]{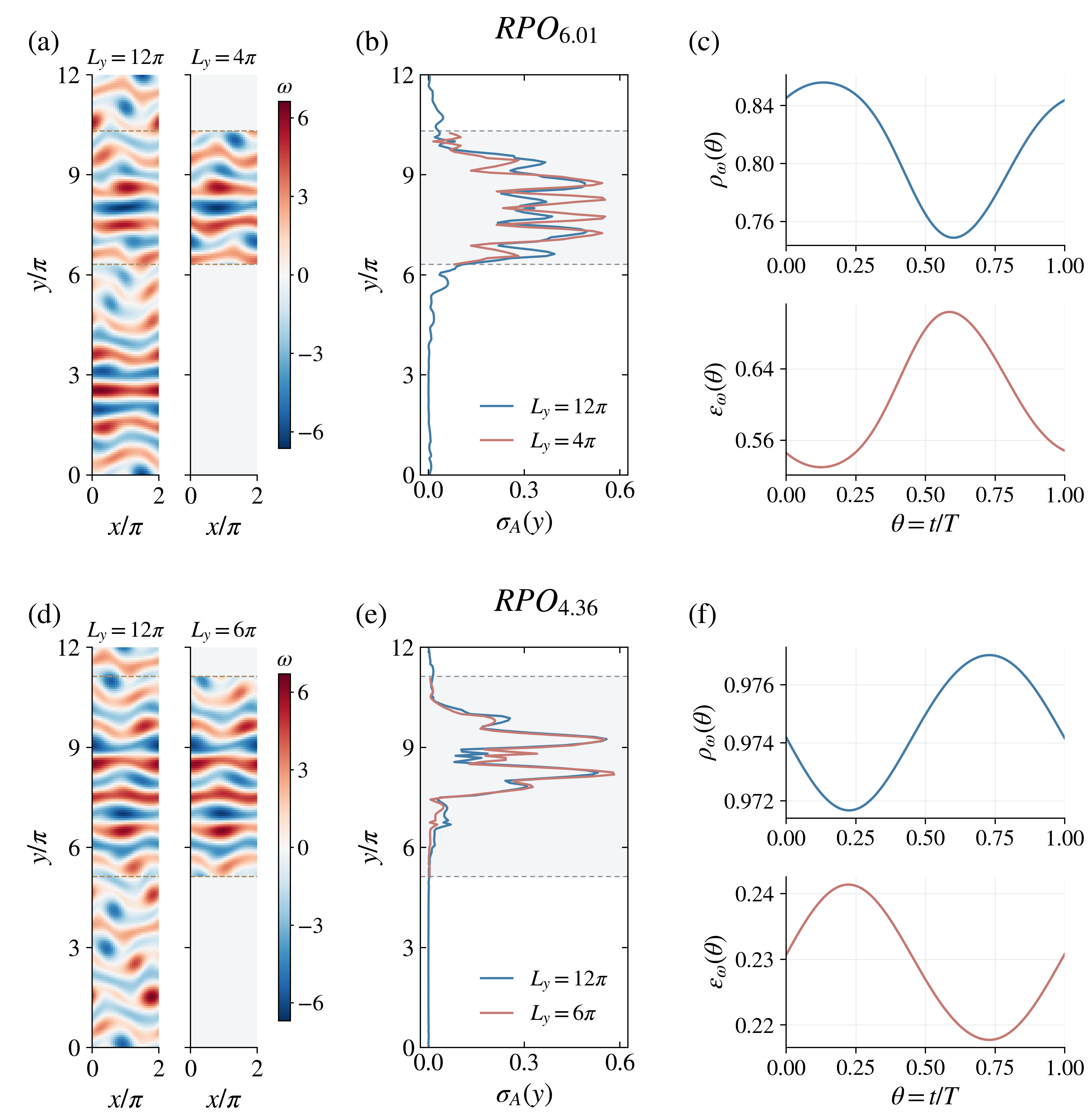}
\caption{
Persistence of localized RPO structure following domain truncation. The upper row (a-c) shows $\mathrm{RPO}_{6.01}$ truncated from $L_y=12\pi$ to $L_y=4\pi$, and the lower row (d-f) shows $\mathrm{RPO}_{4.36}$ truncated from $L_y=12\pi$ to $L_y=6\pi$.
(a,d) Instantaneous vorticity fields of the large and reconverged reduced-domain RPOs. The dashed horizontal lines in the large-domain fields indicate the interval retained to construct the reduced-domain initial condition.
(b,e) Temporal-activity profiles $\sigma_A(y)$ for the large and reduced-domain solutions. The large-domain profile is shown over the full domain, while the reduced-domain profile is positioned over the corresponding retained interval.
(c,f) Phase-resolved spatial Pearson correlation $\rho_\omega(\theta)$ and relative $L^2$ difference $\varepsilon_\omega(\theta)$ between the reduced-domain solution and the corresponding portion of the large RPO, with
$\theta=t/T$. 
}
\label{fig:small_domain_1}
\end{figure*}

To examine whether the localized recurrent dynamics require the full $L_y=12\pi$ cross-stream extent, we construct reduced-domain initial guesses by restricting selected large-domain RPOs to contiguous intervals containing their active cores. The retained intervals and their locations within the large domain are shown explicitly in figure \ref{fig:small_domain_1}. The restricted fields are then mapped onto the corresponding smaller periodic Fourier domains and used solely as initial guesses for independent Newton-Krylov convergence.
Thus, the truncation operation itself does not produce an ECS. Its role is only to provide an initial condition. The resulting state is classified as an RPO only after independent convergence of the smaller-domain Navier-Stokes equations to the prescribed residual tolerance.

Figure~\ref{fig:small_domain_1}a-c illustrate this procedure for $\mathrm{RPO}_{6.01}$. The active core of the $L_y=12\pi$ solution is retained and reconverged in a domain of extent $L_y=4\pi$. The resulting orbit preserves the principal vorticity organization of the large core, although its structure adjusts appreciably after removal of the surrounding flow (see figure~\ref{fig:small_domain_1}a).
This correspondence is quantified using the activity profiles and the phase-resolved distance between the two RPOs. The reduced-domain solution retains the main peaks of $\sigma_A(y)$ found within the corresponding
portion of the large-domain orbit, but the profiles do not coincide pointwise (see figure~\ref{fig:small_domain_1}b).
To quantify this correspondence throughout the recurrent cycle, the two  solutions are compared at equal normalized phase, $\theta=t/T$, after optimization over their relative streamwise translation. 
We 
compute the corresponding Pearson correlation coefficient, i.e.,
$
\rho_\omega(\theta)
=
\left\langle
\omega_{\mathrm{small}}'(\theta),
\omega_{\mathrm{large}}^{\mathrm{loc},\prime}(\theta)
\right\rangle
/
(
\| 
\omega_{\mathrm{small}}'(\theta)
\|_2
\| 
\omega_{\mathrm{large}}^{\mathrm{loc},\prime}
\|_2
), 
$
where primes denote fluctuations about the corresponding spatial means,
and $\omega_{\mathrm{large}}^{\mathrm{loc}}$ denotes the restriction of the large-domain solution to the retained domain.
 For the $L_y=4\pi$ case,
 $\rho_\omega(\theta)\simeq0.75$-$0.86$ and relative $L^2$ distance  $\varepsilon_\omega(\theta)\simeq0.53$-$0.70$ over the cycle (see figure~\ref{fig:small_domain_1}c).
The reduced solution therefore preserves the broad organization of the  active region of the large domain, but undergoes  reorganization
after truncation.

A second truncation provides a considerably stronger correspondence (see figure \ref{fig:small_domain_1}d-f). For this case, the active core is retained in a domain of extent $L_y=6\pi$. The reconverged RPO satisfies $\rho_\omega(\theta)\simeq0.972$-$0.977$ and $\varepsilon_\omega\simeq0.22$-$0.24$ throughout the cycle. 
Thus, both the spatial organization and the phase-dependent evolution of the active region in the large domain  are largely preserved after the inactive surroundings are removed.

Taken together, these truncation tests show that the localized cores of the two large-domain RPOs considered here can persist as exact recurrent solutions after removal of much of the weakly modulated surroundings. This complements the construction of \citet{Zhigunov_Page_2026} from the opposite direction, in which rather than assembling large-domain states from small-domain ECS, we identify localized recurrent structures directly in the extended system and show that their local
recurrent dynamics can be reconverged in smaller domains.

\subsection{Transfer of the learned dynamics across domain sizes}
\label{sec:model_transfer}

\begin{figure}
\centering
\includegraphics[width=0.92\textwidth]{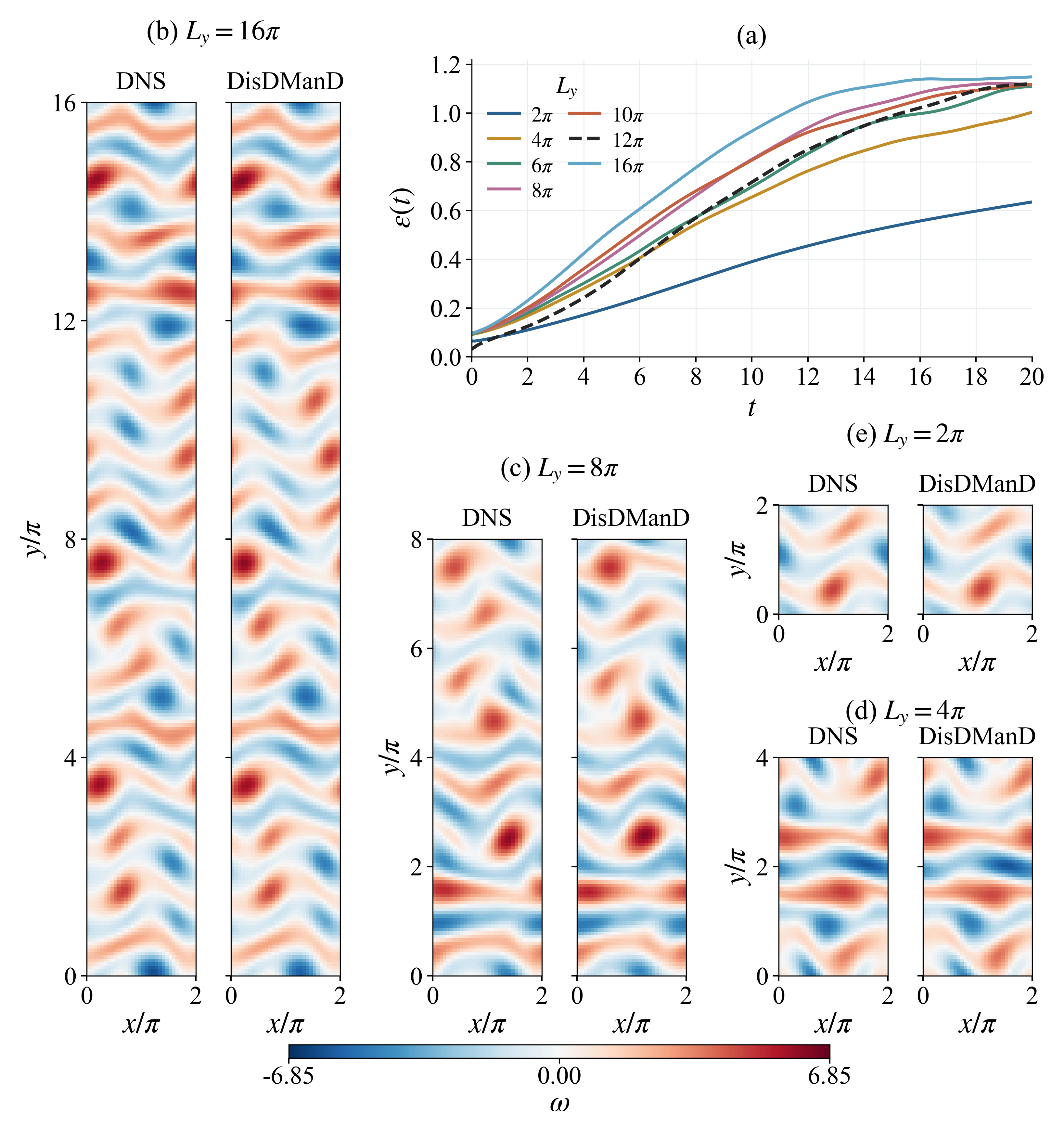}
\caption{
Performance of DisDManD across different $L_y$ domain sizes at
$L_x=2\pi$, $Re=10$ and $n=2$.
(a) Short-time tracking for the DNS and DisDManD averaged over 100 trajectories for different values of $L_y$
(b-e) Representative vorticity fields predicted by DisDManD and DNS at $t=5$ time units, for
$L_y=16\pi$, $8\pi$, $2\pi$, and $4\pi$, respectively.
}
\label{fig:DisDManD_domain_size}
\end{figure}

A consequence of the patch-based formulation is that the learned dynamics are not tied to the global dimension of the training model. The encoder, decoder and latent evolution act locally on spatial patches, while the global latent state is assembled from a variable number of such patches. Changing the cross-stream extent $L_y$ therefore changes the number of local latent variables without altering the learned maps themselves. This differs from  a global reduced-order representation whose input and output dimensions are fixed by the geometry used during training.

We test this property by applying the model trained exclusively in the $[L_x,L_y]=[2\pi,12\pi]$ domain to systems with $L_y=2\pi$, $4\pi$, $6\pi$, $8\pi$, $10\pi$ and $16\pi$, without retraining or fine-tuning. Figure~\ref{fig:DisDManD_domain_size}a shows the short-time tracking as a function of the domain size. 
The initial errors are comparable across the different domain sizes, and the predictions remain fairly accurate over a finite time before separating from the corresponding DNS trajectories.

\begin{table}
\centering
\renewcommand{\arraystretch}{1.15}

\begin{tabular*}{0.96\linewidth}{@{\extracolsep{\fill}}llcccc}
Route & Source & $L_y$ & $T$ & $\ell_x$ & $R$ \\
\hline
Truncation & $\mathrm{RPO}_{6.01}$ & $4\pi$  & 6.38  & -0.10 & $1.95\times10^{-12}$ \\
Truncation & $\mathrm{RPO}_{4.36}$ & $6\pi$  & 3.95  & -0.05 & $2.35\times10^{-11}$ \\
Truncation & $\mathrm{RPO}_{4.36}$ & $6\pi$  & 4.14  & 0.06  & $1.31\times10^{-14}$ \\

Transfer & Independent search & $6\pi$  & 11.55 & -1.29   & $1.20\times10^{-14}$ \\
Transfer & Independent search & $6\pi$  & 12.20 & -0.31  & $2.68\times10^{-14}$ \\
Transfer & Independent search & $6\pi$  & 12.64 & $3.18\times10^{-14}$ & $1.42\times10^{-14}$ \\
Transfer & Independent search & $6\pi$  & 12.69 & -0.04 & $9.39\times10^{-12}$ \\
Transfer & Independent search & $6\pi$  & 12.70 & 0.01 & $2.40\times10^{-14}$ \\
Transfer & Independent search & $6\pi$  & 14.80 & 0.09  & $7.23\times10^{-11}$ \\

Transfer & Independent search & $8\pi$  & 5.03  & -0.67 & $5.08\times10^{-12}$ \\
Transfer & Independent search & $8\pi$  & 17.48 & -2.87  & $5.33\times10^{-12}$ \\

Transfer & Independent search & $16\pi$ & 2.36 & -0.43 &
$6.50\times10^{-4}{}^{*}$ \\
\end{tabular*}

\caption{
Additional RPOs at $Re=10$ obtained through
truncation of existing RPOs and model transfer.
\\
 $^{*}$The $L_y=16\pi$ candidate is included for completeness but is not
classified as an RPO because its residual remains above the convergence tolerance.
}
\label{tab:additional_rpo_routes}
\end{table}

The representative fields in figures~\ref{fig:DisDManD_domain_size}b-e 
compare the vorticity fields between the model and the DNS at $t=5$ time units.
The  model reproduces the principal vorticity structures in systems both smaller and larger than the training domain. In particular, the $L_y=16\pi$  calculation assembles the same learned local maps over eight patches, although only six patches were present during training. At the opposite extreme, for $L_y=2\pi$ the model approaches the travelling-wave state observed in the corresponding DNS. These results indicate that the patch-based representation can be applied to global state dimensions different from that used to train the model.

We next use the transferred model to perform latent multiple-shooting searches and decode the resulting candidates for full-state Newton-Krylov refinement.  As a proof of concept, we
carry out RPO searches at $L_y=6\pi$ and  $8\pi$. In both domains, the transferred model produces initial conditions that subsequently converge to exact RPOs. Table~\ref{tab:additional_rpo_routes} collects the new converged RPOs. At  $L_y=16\pi$,
refinement reduces the residual to $O(10^{-4})$ but does not reach the convergence tolerance,
so this state is reported only as an unconverged candidate.

\begin{figure}
\centering
\includegraphics[width=\textwidth]{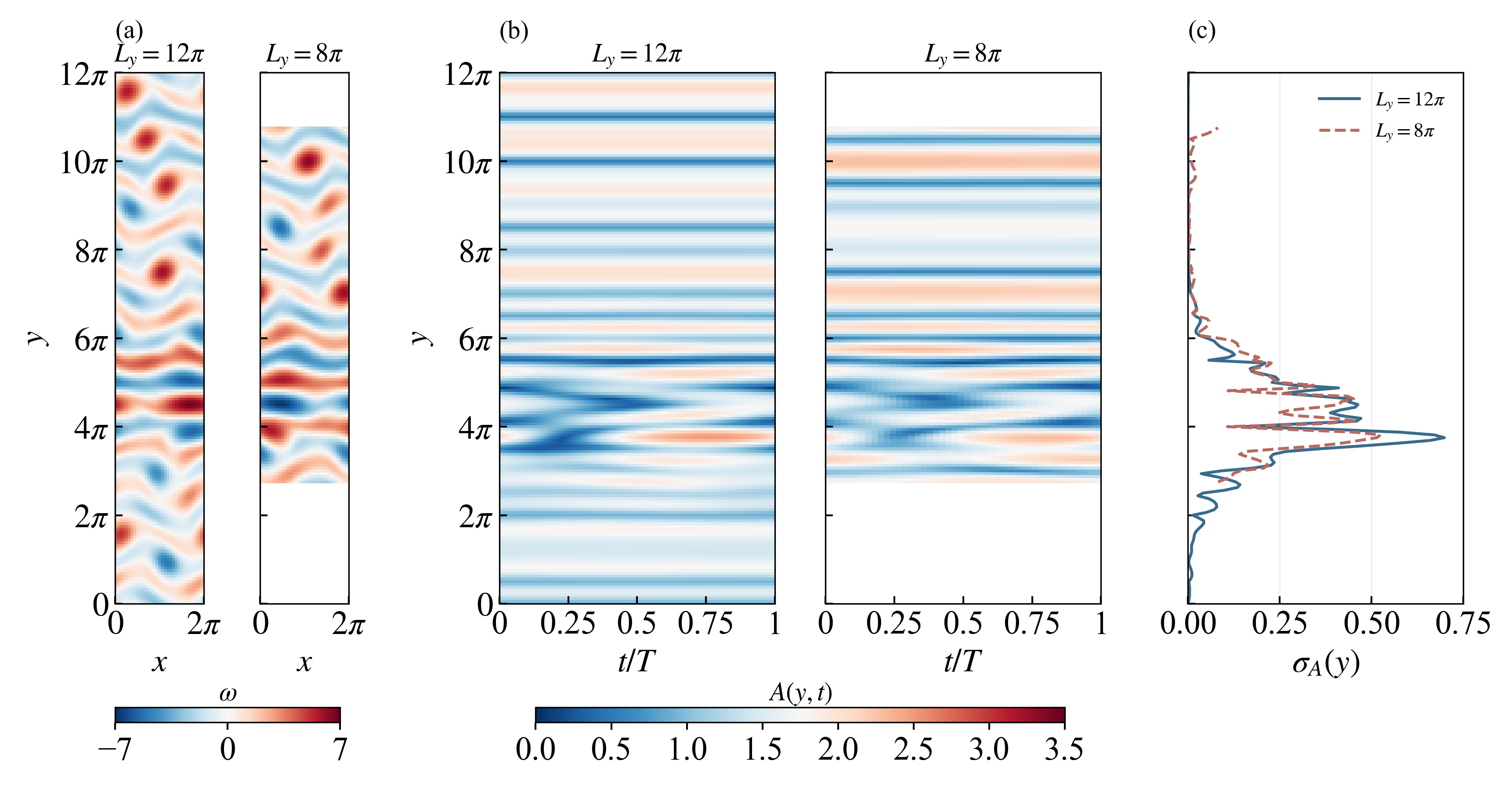}
\caption{ Comparison of localized recurrent dynamics in independently identified RPOs at $L_y=12\pi$ ($\mathrm{RPO}_{4.51}$) and $L_y=8\pi$ ($\mathrm{RPO}_{5.03}$).
(a) Aligned instantaneous vorticity fields.
(b) Corresponding space-time maps of the local fluctuation amplitude $A(y,t)$ over one normalized period.
(c) Temporal-activity profiles $\sigma_A(y)$ for the two RPOs.
}
\label{fig:rpo_domain_comparison}
\end{figure}

A comparison between independently obtained RPOs from Table~\ref{tab:additional_rpo_routes}
provides further evidence that this transfer is associated with  local dynamical structure. Figure~\ref{fig:rpo_domain_comparison} compares
$\mathrm{RPO}_{4.51}$ in the $L_y=12\pi$ domain with $\mathrm{RPO}_{5.03}$
obtained independently in the $L_y=8\pi$ domain. 
The two solutions are aligned in the cross-stream coordinate and temporal phase, with the appropriate symmetry operations also taken into account (see figure~\ref{fig:rpo_domain_comparison}a).
Both RPOs exhibit closely corresponding dynamics within a localized active band. Their space-time fields $A(y,t)$ show similar temporal evolution in this region (see figure~\ref{fig:rpo_domain_comparison}b), while the profiles of
$\sigma_A(y)$ identify active regions of comparable position and extent (see figure~\ref{fig:rpo_domain_comparison}c). The instantaneous vorticity fields provide the corresponding
physical-space picture, in which the organization and evolution of the vortical structures within the active region are largely preserved despite the change in domain size. Outside this region, however, the two solutions can differ substantially.

Taken together with the truncation experiments of section~\ref{sec:reduced_domains}, these results point to a local component of the recurrent dynamics in the extended system. The  selected localized cores can persist as exact RPOs after much of their weakly modulated surroundings is removed. The transfer calculations provide a complementary result: a model trained on local dynamics in one domain can be reused to generate candidates that converge to recurrent solutions in domains of different extent. The comparison between independently obtained RPOs further shows that similar localized recurrent dynamics can occur within distinct global states. These observations do not imply that local RPOs can be combined arbitrarily to form larger-domain solutions; their compatibility with the surrounding flow remains part of the global dynamics. They do, however, support the use of a local representation for both the discovery and interpretation of recurrent dynamics in spatially
extended flow.

\section{Conclusions}
\label{sec:conclusions}

We have developed a data-driven strategy for identifying relative periodic orbits in spatially extended two-dimensional Kolmogorov flow. The central idea is to use a distributed reduced-order model to refine and rank near-recurrent turbulent states before applying the substantially more expensive full-state Newton-Krylov solver. The learned model is used only
to construct initial guesses; all RPOs reported here are subsequently reconverged using the Navier-Stokes equations to the prescribed numerical tolerance. For the $[L_x,L_y]=[2\pi,12\pi]$ domain at $Re=10$, the latent-assisted search yielded fourteen distinct RPOs with periods
$3.65\leq T\leq16.45$, including eight with $T>2\tau_{LE}$. By contrast, none of the ten most promising candidates selected directly from the full-space recurrence search converged under the same full-state refinement procedure. The purpose of the latent model is therefore not to replace the governing
dynamics, but to move turbulent recurrence candidates into more favourable regions of the full-state Newton basins. 
%

The converged solutions reveal substantial variation in their spatial organization. Some RPOs exhibit temporal modulation throughout almost the entire domain, whereas others contain a localized recurrent core surrounded by regions whose vorticity remains finite in amplitude but evolves only weakly over the orbit. Importantly, these weakly modulated surroundings are
not generally close to the laminar state. Spatial localization of recurrence in the present system therefore does not require coexistence between a turbulent recurrent core and a laminar background. Continuation of four
localized branches in $Re$ further shows that the position of the active region can remain approximately fixed while the recurrence period, drift and, in some cases, active extent vary.

Two complementary tests suggest that this localization reflects a genuinely local component of the recurrent dynamics. First, for two selected localized RPOs, retaining the active region and reconverging it in a smaller periodic domain produces exact recurrent solutions of the reduced-domain system. The degree of quantitative correspondence depends on the particular orbit, showing that the surrounding flow can modify the detailed recurrent trajectory even when the localized core persists. Second, because the distributed model acts through shared local maps, the model trained only at $L_y=12\pi$ can be reused without retraining in domains of different extent. Independent searches at $L_y=6\pi$ and $8\pi$ generated candidates that subsequently converged to RPOs under the full-state solver. Together, these observations support a local-to-global interpretation in which recurrent dynamical cores can persist across different global environments, while their compatibility with the surrounding flow remains part of the full system dynamics.
The truncation results provide a complementary perspective to the construction of \citet{Zhigunov_Page_2026}. Rather than assembling an extended-domain solution from prescribed small-domain ECS, we begin with an RPO discovered directly in the extended system, and ask whether its localized active region can persist after the surrounding flow is removed. 

The main result is consequently both computational and dynamical: the
distributed learned dynamics provide an effective route for refining recurrent-state candidates in an extended domain, while the converged Navier-Stokes solutions show that recurrence itself can be strongly localized even when the surrounding flow remains finite in amplitude. These results provide a framework for investigating whether localized invariant solutions form a useful dynamical description of spatiotemporally chaotic flows for which close global recurrence is rare.

Several limitations remain important. The present latent optimization does not include the continuous streamwise shift and therefore enforces periodic, rather than relative-periodic, closure in latent space. The comparison with
direct recurrence is based on a finite candidate set, and the truncation results presently involve only two localized RPOs. Most importantly, the existence of localized recurrent solutions does not by itself establish their dynamical importance to turbulence. Quantifying local turbulent
shadowing, determining their Floquet stability, and characterizing interactions between recurrent cores are therefore natural next steps.
\\

\noindent \textbf{Declaration of Interests}. The authors report no conflict of interest. \\

\noindent\textbf{Data availability.} 
The code and compressed datasets (including videos of the converged RPOs) supporting the findings of this study are publicly available 
at \url{https://github.com/CFTL-Illinois/RPOs_in_spatially_extended_KFlow}.
\\

\noindent\textbf{Acknowledgments.} 
This work used NCSA Delta CPU resources at the National Center for Supercomputing Applications through ACCESS allocations PHY250289 and MCH250041 from the ACCESS
program.

\renewcommand{\theequation}{A\arabic{equation}}
\setcounter{equation}{0}

\section*{Appendix A: Newton-Krylov procedure}
\label{newton_krylov_method}

The full-space Newton-Krylov framework has previously been applied to two-dimensional chaotic falling-film dynamics by \citet{isaac}. We therefore provide only the details required for the present RPO calculations.

Let $G_{T}$ denote the flow map of the discretized Kolmogorov-flow equations and ${S}_{\ell_x}$ a streamwise translation by $\ell_x$. A relative periodic orbit (RPO) satisfies
\begin{equation}
G_{T}(\boldsymbol{\omega}_0)
-
{S}_{\ell_x}\boldsymbol{\omega}_0
=
\boldsymbol{0},
\label{eq:residual_full}
\end{equation}
where $\boldsymbol{\omega}_0$ is the initial vorticity field, $T$ is the period and $\ell_x$ is the streamwise shift.

The RPO is refined using a Jacobian-free Newton-Krylov method \citep{willis2019equilibriaperiodicorbitscomputing}. Defining the unknown vector
\begin{equation}
\boldsymbol{q}
=
(\boldsymbol{\omega}_0,T,\ell_x),
\end{equation}
the RPO condition \eqref{eq:residual_full} is augmented by two phase conditions that remove the neutral directions associated with temporal and streamwise translations. The resulting nonlinear system is denoted by
$\boldsymbol{F}(\boldsymbol{q})=\boldsymbol{0}$. At each Newton iteration, the correction $\delta\boldsymbol{q}$ is obtained from
\begin{equation}
\mathbf{J}\,\delta\boldsymbol{q}
=
-\boldsymbol{F}(\boldsymbol{q}),
\end{equation}
using GMRES. Jacobian-vector products are evaluated in matrix-free form as
\begin{equation}
\mathbf{J}\,\delta\boldsymbol{q}
\approx
\frac{
\boldsymbol{F}(\boldsymbol{q}
+\varepsilon\,\delta\boldsymbol{q})
-
\boldsymbol{F}(\boldsymbol{q})
}{\varepsilon},
\end{equation}
thereby avoiding explicit construction of the Jacobian.

The Newton iteration was considered converged when the Euclidean norm of the augmented residual satisfied
$\|\boldsymbol{F}\|_2\leq10^{-8}$.
Each refinement was allowed at most 100 Newton iterations. The GMRES Krylov dimension was initialized at 20 and increased when required to achieve sufficient reduction of the linear residual. Globalization was achieved using backtracking with trial step lengths
$\alpha=2^{-j}$, $j=0,\ldots,8$. A trial step was accepted only if the updated period remained within the prescribed bounds and the nonlinear residual was finite and smaller than that of the current iterate. A search was terminated unsuccessfully if the Newton-iteration limit was reached or if no admissible residual-decreasing step could be found.

\section*{Appendix B: Neural-networks details
\label{Appendix_NN}}

\begin{figure}
    \centering
 \includegraphics[width=\linewidth]{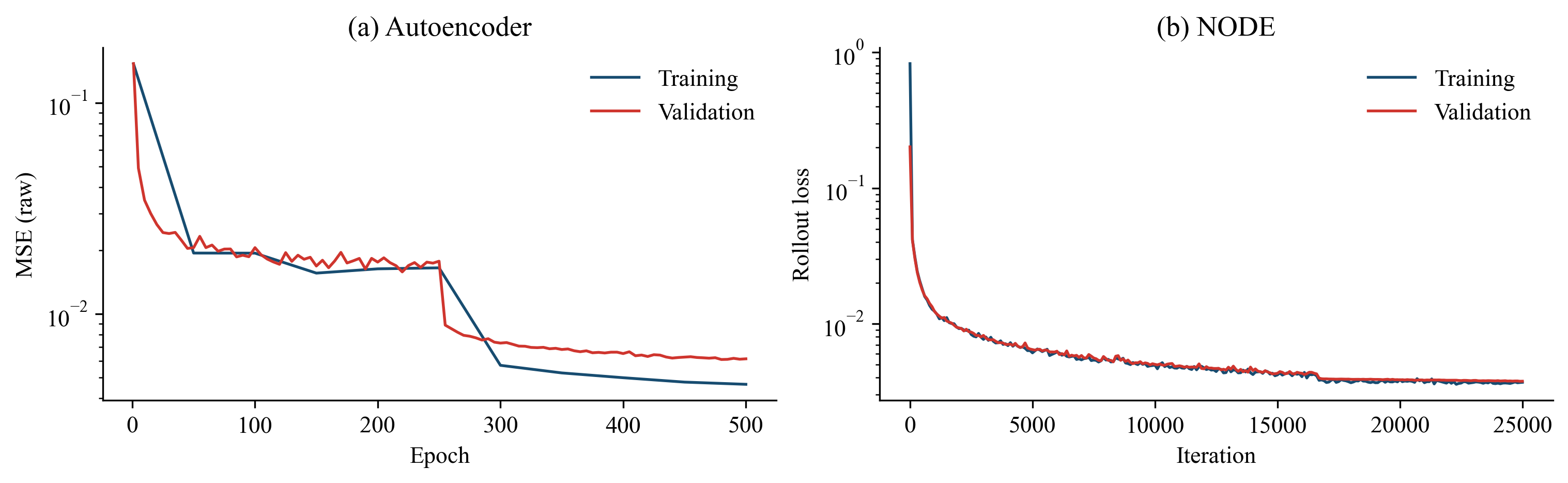}
\caption{Training histories for the neural-network components of
DisDManD. (a) Mean-squared reconstruction error of the autoencoder for
the training and validation datasets. (b) Latent-space trajectory rollout loss of the NODE for the training and validation datasets.}
\label{fig:network_training}
\end{figure}

This appendix provides the architecture and training details of the autoencoder and neural ordinary differential equation (NODE) used in the DisDManD model. The corresponding training and validation losses are
shown in figure~\ref{fig:network_training}.

For $Re=10$, in a domain of $[L_x,L_y]=[2\pi,12\pi]$, the training dataset was generated from 25 independent trajectories. Each trajectory was initialized from a different seed, with a distinct perturbation applied to its initial condition in order to sample different
regions of the chaotic attractor. The flow fields were sampled at a fixed interval of $\Delta t=0.1$. In total, the simulations produced 218\,031 flow-field snapshots, which were subsequently used to train
the data-driven model.

The autoencoder thus provides a dimension reduction of $\mathbb{R}^{6144}\longrightarrow\mathbb{R}^{120}$ (i.e., $d_h=20$ latent variables per patch).
We overlap patches with $N_{ov}=1$ in this work  to mitigate sharp changes at the domain boundaries.
The encoder and decoder each contain two hidden layers of widths 5000 and 2000, with sigmoid activation functions. 
The autoencoder is trained for 500 epochs by minimizing the mean-squared reconstruction error between the input vorticity fields and their reconstructions. The learning rate is initially set to $10^{-3}$ and reduced by one order of magnitude, to $10^{-4}$, after 250 epochs. As shown in figure~\ref{fig:network_training}a, the training and validation errors remain comparable throughout training, with no appreciable separation after convergence.

The latent-space dynamics are represented using a locally coupled NODE. For each spatial patch $k$, the NODE predicts the evolution of its local latent state $\boldsymbol{h}^k\in\mathbb{R}^{20}$ using both $\boldsymbol{h}^k$ and the latent states of its two neighbouring patches. Thus, the input to each local NODE has dimension $3d_h=60$, whereas its output has dimension $d_h=20$. 
Although the time derivative of each patch is predicted from only that patch and its immediate neighbours, the six coupled equations are integrated simultaneously to evolve the complete latent state
$\boldsymbol{h}(t)\in\mathbb{R}^{120}$. Consequently, information can propagate between patches throughout a rollout, and the predicted state at each snapshot time depends recursively on the preceding globally coupled evolution.

The NODE contains four hidden layers, each of width 400. The first two hidden layers use ReLU activation functions, whereas the final two use sigmoid activations. Training is performed for 25000 iterations using a batch
size of 256 and an initial learning rate of $10^{-3}$. Each training sample comprises a window of 10 consecutive snapshots. Starting from the encoded state at the beginning of the window, the coupled NODE system is integrated over the complete window. The predicted latent trajectory is then compared with the corresponding encoded DNS trajectory across all six patches. The training
and validation rollout losses are shown in figure~\ref{fig:network_training}b.

\section*{Appendix C: Distance of inactive regions from the laminar state}
\label{app:inactive_regions}

\begin{figure}
    \centering
 \includegraphics[width=\linewidth]{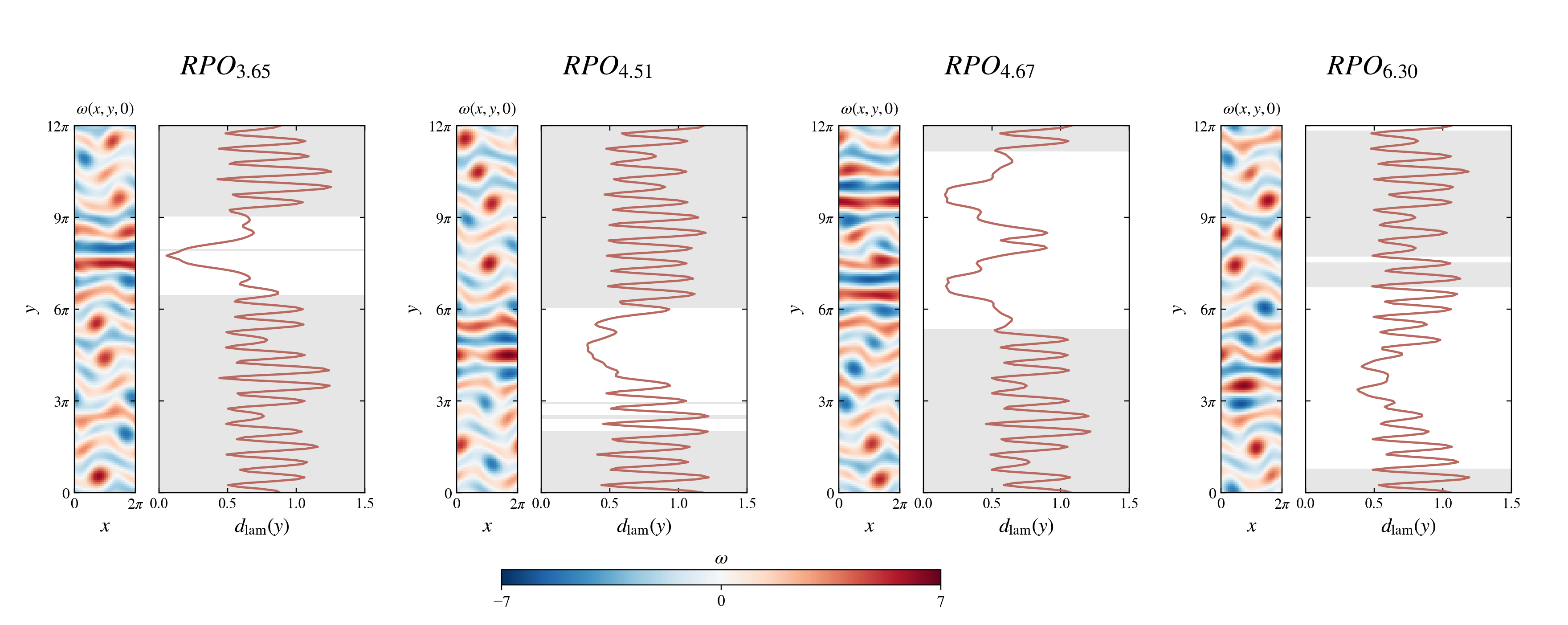}
\caption{
Spatial organization of the remaining localized RPOs. For each orbit, the left panel shows the instantaneous vorticity field
$\omega(x,y,0)$ and the right panel shows the corresponding distance from
the laminar solution, $d_{\rm lam}(y)$, computed over one RPO period.
Shaded regions denote locations classified as inactive.
}
\label{RPO_dlam_appendix}
\end{figure}

Figure~\ref{RPO_dlam_appendix} extends the analysis of
\S~\ref{sec:localization} to the remaining RPOs possessing substantial inactive regions. For each orbit, we show an instantaneous vorticity field together with the local distance from the laminar solution,
$d_{\rm lam}(y)$, defined in equation~\eqref{eq:dlam}. The shaded intervals
identify inactive regions according to the same activity threshold
$\sigma_A(y)\leq\sigma_c$ used in the main text.
Although $d_{\rm lam}(y)$ may exhibit local minima within some inactive intervals, it remains finite throughout most of these regions. The weakly modulated surroundings of the localized RPOs of this work do not correspond to laminar patches.

\bibliographystyle{jfm}
\bibliography{references}

\end{document}